\def\aj{AJ}%
\def\apj{ApJ}%
\def\apjl{ApJ}%
\def\apjs{ApJS}%
\def\aap{A\&A}%
\def\aaps{A\&AS}%
\def\mnras{MNRAS}%
\def\pasp{PASP}%
\def\pasj{PASJ}%
\def\physrep{Phys.~Rep.}%
\newcommand{\ec}{\mbox{$E_{\rm 0}$}}
\newcommand{\kms}{\mbox{km$\,$s$^{-1}$}}
\newcommand{\kmspkpc}{\mbox{km$\,$s$^{-1}$\,${\rm kpc}^{-1}$}}
\newcommand{\msun}{\mbox{${\rm M}_{\odot}$}}

\newcommand{\tsh}{\mbox{$t_{\rm ds}$}}
\newcommand{\tdis}{\mbox{$t_{\rm dis}$}}
\newcommand{\tdrift}{\mbox{$t_{\rm drift}$}}
\newcommand{\vdrift}{\mbox{$V_{\rm drift}$}}

\newcommand{\Mc}{\mbox{$M_{\rm c}$}}
\newcommand{\Mcl}{\mbox{$M_{\rm c}$}}

\newcommand{\rh}{\mbox{$r_{\rm h}$}}
\newcommand{\rhsq}{\mbox{$r^2_{\rm h}$}}
\newcommand{\rhcube}{\mbox{$r^3_{\rm h}$}}
\newcommand{\rrms}{\mbox{$\bar{r^2}$}}
\newcommand{\yrms}{\mbox{$\bar{y^2}$}}
\newcommand{\rt}{\mbox{$r_{\rm t}$}}
\newcommand{\rc}{\mbox{$r_{\rm c}$}}
\newcommand{\rv}{\mbox{$r_{\rm v}$}}

\newcommand{\rcr}{\mbox{$R_{\rm CR}$}}
\newcommand{\gm}{\mbox{$g_{\rm m}$}}
\newcommand{\gmsq}{\mbox{$g^2_{\rm m}$}}
\newcommand{\width}{\mbox{$\Delta\,\phi_{1/2}$}}
\newcommand{\widthkpc}{\mbox{$W_{1/2}$}}

\newcommand{\vdisc}{\mbox{$V_{\rm disc}$}}
\newcommand{\dr}{\mbox{${\rm d}$}}
\newcommand{\erf}{\mbox{${\rm erf}$}}
\newcommand{\Yc}{\mbox{$Y_{\rm c}$}}
\newcommand{\vy}{\mbox{$v_{\rm y}$}}
\newcommand{\vmax}{\mbox{$V_{\rm max}$}}
\newcommand{\vmaxsq}{\mbox{$V^2_{\rm max}$}}

\newcommand{\deimp}{\mbox{$\Delta E_{\rm imp}$}}
\newcommand{\de}{\mbox{$\Delta E/|E_0|$}}
\newcommand{\dm}{\mbox{$\Delta M/M_0$}}
\newcommand{\tcr}{\mbox{$t_{\rm cr}$}}
\newcommand{\Tcr}{\mbox{$T_{\rm cr}$}}

\newcommand{\sigmah}{\mbox{$\Sigma_{\rm HI}$}}

\newcommand{\meanrhoh}{\mbox{$\langle\rho_{\rm HI}\rangle$}}
\newcommand{\Hz}{\mbox{$H_{\rm z}$}}
\newcommand{\As}{\mbox{$A_{\rm s}(x)$}}
\newcommand{\Aw}{\mbox{$A_{\rm w}(x)$}}
\newcommand{\msunpccube}{\mbox{$M_{\odot}\,{\rm pc}^{-3}$}}
\newcommand{\omegap}{\mbox{$\Omega_{\rm p}$}}

\documentclass[useAMS,usenatbib]{mn2e}
\usepackage{amssymb,latexsym,graphicx,natbib}

\title[The effect of spiral arm passages on the evolution of stellar clusters]
  {The effect of spiral arm passages on the evolution of stellar clusters}

\author[M. Gieles et al.]
  {M.~Gieles,$^{1,2,3}$ E.~Athanassoula,$^2$ and S.F.~Portegies Zwart,$^3$\\
  $^1$Astronomical Institute, Utrecht University, 
  Princetonplein 5, 3584 CC Utrecht, The Netherlands \\
  $^2$ LAM, Observatoire Astronomique de Marseille Provence, 
  2 Place le Verrier, 13248 Marseille Cedex 4, France\\ 
  $^3$ Astronomical Institute `Anton Pannekoek', University of Amsterdam, 
  Kruislaan 403, 1098 SJ Amsterdam, The Netherlands, \\
  Section Computational Science, 
  University of Amsterdam, Kruislaan 403, 1098 SJ, The Netherlands}

\date{Released 2006 Xxxxx XX}

\pagerange{\pageref{firstpage}--\pageref{lastpage}} \pubyear{2006}

\def\LaTeX{L\kern-.36em\raise.3ex\hbox{a}\kern-.15em
    T\kern-.1667em\lower.7ex\hbox{E}\kern-.125emX}

\begin{document}         
\maketitle

   \begin{abstract} We study the effect of spiral arm passages on the
evolution of star clusters on planar and circular orbits around the
centres of galaxies. Individual passages with different relative
velocity (\vdrift) and arm width are studied using $N$-body
simulations. When the ratio of the time it takes the cluster to cross
the density wave to the crossing time of stars in the cluster is much
smaller than one, the energy gain of stars can be predicted accurately
in the impulsive approximation. When this ratio is much larger than one,
the cluster is heated adiabatically and the net effect of heating is
largely damped. For a given duration of the perturbation, this ratio is
smaller for stars in the outer parts of the cluster compared to stars
in the inner part. 
The cluster energy gain due to perturbations of various duration as
obtained from our $N$-body simulations
is in good agreement with theoretical predictions taking into account
the effect of adiabatic damping.
Perturbations by the broad stellar component of the spiral arms on a
cluster are in the adiabatic regime and, therefore, hardly contribute
to the energy gain and mass loss of the cluster. We consider the
effect of crossings through the high density shocked gas in the spiral
arms, which result in a more impulsive compression of the cluster.
The energy injected during each spiral arm passage can exceed the
total binding energy of a cluster, but, since most of the energy goes
in high velocity escapers from the cluster halo, only relatively
little mass is lost.  A single perturbation that injects the same
amount of energy as the initial (negative) cluster energy causes at
most $\sim$40\% of the stars to escape. We find that a perturbation
that delivers $\sim$10 times the initial cluster energy is needed to
completely unbind a cluster in a single passage.
The net effect of spiral arm perturbations on the evolution of low
mass ($\lesssim100\,\msun$) star clusters is more profound than on
more massive clusters ($\gtrsim1000\,\msun$). This is due to the fact
that the 
crossing time of stars in the latter is shorter, causing a larger
fraction of stars to be in the adiabatic regime. The time scale of
disruption by subsequent spiral arm perturbations depends strongly on
position with respect to the radius of corotation (\rcr), where
$\vdrift=0$. The time between successive encounters scales with
\vdrift\ as $\tdrift\propto\vdrift^{-1}$ and the energy gain per passage
scales as $\Delta E\propto V^2_{\rm drift}$. Exactly at \rcr\ passages
do not 
occur, so the time scale of disruption is infinite. The time scale of
disruption is shortest at $\sim$0.8-0.9$\,\rcr$, since there \vdrift\
is low.  This location can be applicable to the solar neighbourhood. In
addition, the four-armed spiral pattern of the Milky Way makes spiral
arms contribute more to the disruption of clusters than in a similar
but two-armed galaxy. Still, the disruption time due to spiral arm
perturbations there is about an order of magnitude higher than what is
observed 
for the solar neighbourhood, making spiral arm perturbations a
moderate contributor to the dissolution of open clusters. 

\end{abstract}
\begin{keywords}
methods: $N$-body simulations -- galaxies: star clusters, spiral -- Galaxy: open clusters and associations: general
\end{keywords}

\section{Introduction}
\label{sec:introduction7}

The distinct difference between open and
globular clusters has vanished since the discovery of young massive
clusters in merging and interacting galaxies
(e.g. \citealt{1992AJ....103..691Hmnras};
\citealt{1999AJ....118.1551W}). There is, however, still an evident difference in
evolution. Star clusters formed in discs, which in our Galaxy are
referred to as open clusters, experience external perturbations by
giant molecular clouds (GMCs) and by spiral arms and other disc
density perturbations. These are not present in the halo of a galaxy,
where most of the globular clusters reside. To understand how cluster
populations, such as the open clusters in the solar neighbourhood
\citep{2005A&A...438.1163K} and such as the ones found in spiral
galaxies like M51
\citep{2005A&A...431..905B} and NGC~6946
\citep{2001ApJ...556..801L}, evolve, it is important to understand the
effect of these external perturbations.

The number of Galactic open clusters as a function of age shows a lack
of old open clusters, first pointed out by
\citet{1958RA......5..507O}. This lack can partially be explained by
the rapid fading of clusters with age due to stellar evolution, which
makes it harder to observe them at older ages. Still, fading can not
explain the difference between the observed and the expected number of
old ($\gtrsim 1\,$Gyr) open clusters, implying that a significant
fraction must have been destroyed (\citealt{1971A&A....13..309W};
\citealt{2005A&A...441..117L}). In addition to two-body relaxation,
clusters in tidal fields dissolve by the combined effect of: A.) Mass
loss due to stellar evolution, reducing the mass and hence the binding
energy of the cluster; B.) the tidal field of the host galaxy,
imposing a tidal boundary, increasing the escape rate of stars; C.)
bulge/disc shocks, pushing stars over the tidal boundary and D.)
additional perturbations induced by irregularities in the galaxy,
such as GMCs and spiral arms. The importance of the first three
effects has been studied in detail by many people
(for example \citealt{1990ApJ...351..121C}; \citealt{1997ApJ...474..223G};
\citealt*{1999ApJ...522..935G}; \citealt{2000ApJ...535..759T};
\citealt{2003MNRAS.340..227B}), using different techniques
(e.g. Fokker-Planck calculations or $N$-body simulations).

These studies were mainly aimed at understanding the evolution of
globular clusters residing in the Galactic halo. 
The observed short disruption time
($\sim100\,$Myr) of disc clusters in the grand-design spiral galaxy M51
\citep{2005A&A...441..949G} is an order of magnitude lower than
expected from the tidal field of the galaxy
\citep*{2005A&A...429..173L}. Even for clusters in the solar
neighbourhood, $N$-body simulations predict disruption times five
times longer than observed. These $N$-body simulations, however,
ignore the presence of spiral arms (for example
\citealt{2000ApJ...535..759T} and \citealt{2003MNRAS.340..227B}). In
this paper, we study the contribution of spiral density waves to the
cluster dissolution time.  

Spiral arms are believed to be density waves rotating around the
galaxy centre with an angular pattern speed (\omegap)  
which is independent of the distance $R$ to the galactic
centre (see \citealt{1984PhR...114..321A} for a review). 
The radius at which \omegap\ is equal to the angular velocity in the
disc ($\Omega (r) = \vdisc/R$, where \vdisc\ the circular velocity in
the disc) is called the corotation radius ($\rcr$). 
As gas travels through such a density wave,
it gets shocked and compressed to five to ten times higher densities
\citep{1969ApJ...158..123R}. This is observable as narrow sharp dust
lanes in optical images of (nearly) face-on spiral galaxies which are
far from edge-on.

A cluster on a circular orbit around the centre of the galaxy will, in
the inertial frame, have a velocity equal to the circular velocity
\vdisc. In the reference frame of the spiral arm, its velocity
(hereafter drift velocity, $\vdrift$) is equal to

\begin{eqnarray}
\vdrift & = & \vdisc-\omegap R \nonumber\\
        & = & \vdisc\left[1-R/\rcr\right].
\label{eq:vdrift}
\end{eqnarray}

 The absolute value of $\vdrift$ decreases going from the galaxy centre
to \rcr, where it is zero, and increases again beyond that radius. The
time it takes for a cluster to travel from one arm to the next is
defined as the drift time: $\tdrift(R)\equiv2\pi R/(m\,|\vdrift(R)|)$,
where $m$ is the number of spiral arms in the galaxy.  Thus, near to
\rcr, a cluster experiences few passages but of long duration, while
at considerably smaller, or considerably larger radii it undergoes
many short-lasting spiral arm passages with high velocity. As it moves
from the low density inter-arm region to the high density in the arms,
the cluster gets compressed due to tidal forces. These accelerate the
cluster stars, some of which may reach the escape velocity and become
unbound.

This is comparable to what happens to a globular cluster that crosses
the Galactic disc \citep*{1972ApJ...176L..51O}.  There are, however,
important qualitative differences between spiral arm perturbations and
disc shocks. In particular, globular clusters cross the disc with a
velocity almost independent of $R$. On the contrary, the velocity of
the cluster with respect to the spiral depends strongly on the
distance to the centre.  Similarly, the scale-height of a disc is
independent of $R$, provided the galaxy is not barred
\citep{1981A&A....95..105V, 1981A&A....95..116V, 1982A&A...110...61V,
1982A&A...110...79V, 1996A&AS..117...19D}. For spiral arms, on the
contrary, it is the angular width of the arm that is roughly
independent of radius, so that the linear width of a spiral arm is
proportional to the radius.

The passage duration determines the nature of the perturbation of the
arm on the cluster. If the time to cross the arm (\Tcr) is much longer
than the crossing time of stars in the cluster (\tcr), the stars can
adiabatically adjust to the density increase. When \Tcr\ is much
shorter than \tcr, the stars get an impulsive velocity increase due to
tidal forces. Stars in the cluster core have a short \tcr\ and,
therefore, get adiabatically heated. The boundary between adiabatic
and impulsive heating in the cluster is determined by the ratio
\tcr/\Tcr. If this ratio is low, a higher fraction of stars is heated
impulsively.

In this paper we study the effect of spiral arm passages on clusters
and investigate the importance of relative velocity with respect to
the arm. This paper is organised as follows: In
Section~\ref{sec:parameters} the physical parameters of spiral arms
are derived from observations. Simple analytical estimates of the
effects of compressive perturbations are given in
Section~\ref{sec:theory}. They are confronted with results of $N$-body
simulations in Section~\ref{sec7:nbody}. In
Section~\ref{sec7:disruption} the results of the simulations are used
to derive cluster mass loss and disruption times due to the spiral arm
perturbations. The conclusions are presented in
Section~\ref{sec:conclusion}.

 
\section{Spiral arm parameters}
\label{sec:parameters}

\subsection{Stellar arms}

Disc clusters are found to have short disruption
times both in our Galaxy and in M51. Therefore, we will hereafter
consider two fiducial spiral galaxies, one Milky-Way-like -- which we
will call for brevity Milky Way, or MW -- and the other M51-like --
which we will call grand design spiral.  We will adopt for both a
constant rotational velocity of $\vdisc=220\,\kms$, that is, we limit
ourselves to the regions where their respective rotation curves are
flat.

Spiral patterns are present in both young and old populations,
as already noted by \citet{1955PASP...67..232Z} and later by
\citet{1976ApJS...31..313S}. The old Population II stars in the disc
show a broad spiral pattern. In contrary, the young Population I
stars, which form preferentially in the spiral arms, show more
irregular spiral patterns.  Estimates of the arm-interarm density
contrast ($\Gamma$) of the stellar population in the disc of M51 were
made by
\citet{1993ApJ...418..123R} and by
\citet{1996ApJ...460..651G}, using the $K$-band light as
tracer of stellar mass. Their results are in good agreement and show
that the maximum value for $\Gamma$ is around $2-3$. Similar values
have been found for other grand design spirals (e.g. NGC~4254, see
\citealt*{2001ApJ...562..164K}). We expect our own Galaxy to
have smaller values of $\Gamma$. Indeed, \citet{2001ApJ...556..181D}
find a value of $\Gamma\simeq1.3$ from model comparisons to $K$-band
measurements.

M51 has a clear two-armed ($m$ = 2) grand-design spiral, presumably
due to the interaction with its companion NGC~5195. The structure of
our Galaxy is more complex. It has a bar (see
\citealt{2002ASPC..275..105D} for a review and references therein)
 and a spiral structure which could rotate with a pattern speed
different from that of the bar \citep{1987ApJ...318L..43T,
1988MNRAS.232..733S}. \citet{2000A&A...358L..13D} and
\citet{2001ApJ...556..181D}
modelled the COBE-DIRBE data and propose that the old stellar
population has a two armed spiral, while the gas and young and stars
form a four-armed spiral (for the latter see also
\citealt{1976A&A....49...57G}). Studies in which all components (bar
and spirals) have the same pattern speed find that the solar
neighbourhood is well beyond corotation \citep{2000AJ....119..800D,
2001A&A...373..511F}, but models allowing for a different pattern
speed for the bar and spiral components give a better view of the
complex structure in our Galaxy. \citet{2003MNRAS.340..949B}
calculate the gas response to a composite bar plus spiral model and
find best agreement for the observations with a pattern speed of
$60\,\kms\,$kpc$^{-1}$ for the bar and $20\,\kms\,$kpc$^{-1}$ for the
spiral.  Similar values are found by
\citet{2004MNRAS.350L..47M}. This places the solar neighbourhood well
beyond the bar corotation, but just within the spiral corotation. 
With an adopted solar radius of $R_0=8.5\,$kpc and the adopted value
for $\vdisc=220\,\kms$ the solar neighbourhood is at $R\simeq0.8\,\rcr$. Compared to the
Milky Way, M51 has been modelled very little and no clear-cut value
has been given for its pattern speed.  Recently,
\citet*{2004ApJ...607..285Z} and
\citet*{2004PASJ...56L..45E} determined the pattern speed of four
grand design two-armed spirals and they find values, between 30 and 40
$\kms\,$kpc$^{-1}$.  We adopt \rcr= 6 kpc, resulting in a pattern
speed of $\sim 37\,\kmspkpc$.

To further quantify the effect of a spiral-arm passages on the
evolution of star clusters, we also need to consider the width of the
arms.  To enable a quantitative discussion about spiral-arm width, we
define $\width$ as the arm annular full width at half maximum.
The typical values of \width\, measured from optical images of six
spiral galaxies by \citet{1976ApJS...31..313S}, is almost independent
of $R$ and $\width \simeq 20^\circ$. \citet{1998MNRAS.299..685S}
studied $K$-band images of a sample of 45 face-on spiral galaxies and
found for two-armed spiral galaxies typical values of $\width$ between
$20^{\circ}$ and $40^{\circ}$. These values are slightly higher than
found by
\citet{1976ApJS...31..313S}, probably because the $K$-band is a
better tracer of the old stellar populations, for which \width\ is
larger. For $\width=30^\circ$, the value of \Tcr\ depends on \vdrift\
and $R$ as $\Tcr\simeq0.5R/|\vdrift|$. For the solar neighbourhood
$\Tcr\simeq80\,$Myr. For our M51-like galaxy, the values for \Tcr\ at
$[0.3/0.5]\times\rcr$ are $[14,68]\,$Myr. The values for \Tcr\
are all higher than the typical \tcr\ of clusters, that is, a few
Myrs.  Therefore, the stars in the cluster respond adiabatically
when the cluster crosses the arm and this crossing does not have the short lasting
compressive effect as a galactic disc has on globular clusters. 


\subsection{Gaseous arms}
\label{subsec:gaseous}
The gaseous spiral pattern is different from the stellar
one. Gas gets shocked in the potential well of the stellar arms, where
it gets compressed to up to ten times higher densities
\citep{1969ApJ...158..123R}. These high gas densities manifest
themselves in optical observations of spiral galaxies as thin dust
lanes on the leading side of the arm within \rcr.  After the shock,
the density decreases exponential-like  to a value slightly lower
than the mean gas density, as is illustrated in
Fig.~\ref{fig:peak}. Such strongly peaked high gas densities act as a
compressive perturbation on the cluster (Section~\ref{sec:theory}),
which can be compared to disc shocks of globular clusters. The disc
crossing is referred to as a shock in that context because of the
short duration of the perturbation. That is, the time to cross the
disc is much shorter than \tcr. Since we here also discuss shocks by
gas in the arm we avoid confusions and refer to perturbations instead
of shocks. In addition, in Section~\ref{sec:theory} we will show that,
due to the variation of
\Tcr\ with $R$, the perturbations are of highly varying
duration. 

Besides the higher densities, also the typical values of
\width\ of the gaseous spiral arm are much smaller than the one of the
stars. \citet{1969ApJ...158..123R} argues that the width can be
calculated by the time scale of formation and evolution of massive
stars, that is, a few tens of Myr, since these stars are visible in
the optical, so they must have traveled out of the high density
dusty environment where they formed.  Studies of the molecular spiral arms
in M31 and M51 by \citet{1994PASJ...46..527N} and
\citet{2006A&A...453..459N}, give values of the full-width at half
maximum (\widthkpc) of $500\,$pc and $1000\,$pc, respectively. Since
these values are close to the resolution of the beam, the sharp peak
in density (Fig.~\ref{fig:peak}) is presumably not resolved.
Hydrodynamical simulations of barred galaxies revealed narrower density
peaks of the gas (see Fig.~11 of \citealt{1992MNRAS.259..345A}), very
close to what was predicted by \citet{1969ApJ...158..123R} and what is
shown in Fig.~\ref{fig:peak}. We also measured the width of the dust
lane on an optical picture of M51 and the width of the HI on the 8
arcsec resolution map of \citet{1990AJ....100..387R} and find that
they are both compatible with a width of 250 pc. We, therefore, adopt
$\widthkpc=250\,$pc for the gaseous arms.

Measurements of the neutral hydrogen content in a sample of spiral
galaxies in the Virgo cluster \citep{1994AJ....107.1003C} show that
the {\it surface} density of HI gas ($\sigmah$) within $R\simeq10\,$kpc in
gas rich spirals is nearly constant with $R$ and of the order of $\sigmah\simeq
10\,\msun{\rm pc}^{-2}$. A recent study of the Milky Way
\citep*{2006Sci...312.1773L} gives similar values for $\sigmah$ in the
solar neighbourhood. \citet{2006Sci...312.1773L} also find that
$\Gamma$ is almost independent of $R$. Their measurements of the
arm-interarm surface density contrast, combined with the measured
vertical compression factor in the spiral arms are consistent with
$\Gamma=10$ for the gas.

From \sigmah\ and the vertical scale height of gas ($\Hz$) we can
derive the mean midplane density of HI (\meanrhoh), since
$\meanrhoh=\sigmah/2\Hz$, assuming an exponential vertical density
distribution. \citet{1988A&A...192..117V} finds a vertical scale
height of $\Hz\simeq150\,$pc, resulting in
$\meanrhoh\simeq0.033\,\msun\,$pc$^{-3}$. The peak density ($\rho_0$)
is then

\begin{equation}
\rho_0\simeq\Gamma\,\meanrhoh,
\label{eq:rho0}
\end{equation}
where we assume $\Gamma$ is the ratio of $\rho_0$ to
$\meanrhoh$.

All parameters of spiral arms, used in our models, are summarised in
 Table~\ref{tab:parameters}.

\begin{table}
\caption{Adopted parameters for spiral arms in our fiducial grand design two-armed spiral
 galaxy and in the Milky Way}
\label{tab:parameters}                                              
\begin{center}
\begin{tabular}{lrrrr}
\hline 
                           & Grand design       & Milky Way     \\\hline 
$m$                        & 2                  & 4             \\
$\vdisc\,(\kms)$           & 220                & 220           \\
$\Omega_p\,(\kmspkpc)$     & 36.7               & 24.5          \\          
$\rcr\,$(kpc)              & 6                  & 9             \\
$\width$(stars)            & $30^{\circ}$       & $15^{\circ}$  \\
$\widthkpc$(gas) (pc)      & 250                & 250           \\
$\Gamma_{\rm stars}$       & 2-3                & $<2$          \\
$\Gamma_{\rm gas}$         & 10                 & 10            \\
$\meanrhoh$ ($\msun\,$pc$^{-3}$) & 0.033		& 0.033		\\
\hline
\end{tabular}
\end{center}

\end{table}

\subsection{Potential-density relations for the arms}
\label{subsec:potdens}
\begin{figure}
\centering
   \includegraphics[width=8.cm]{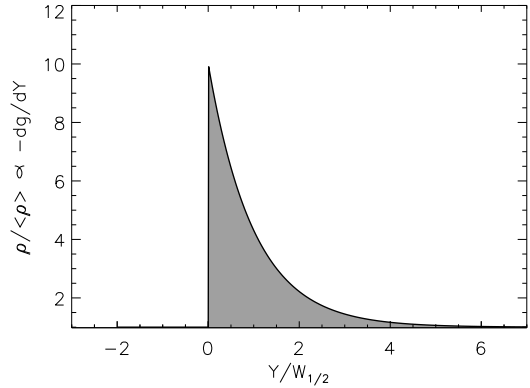}
   
    \caption{Schematic illustration of the gas density along azimuthal
    trajectories through a spiral arm, where the gas travels from left
    to right in this figure. The grey shaded area is the gas
    density predicted by \citet{1969ApJ...158..123R}. 
    When the scale length of
    a Gaussian function ($H$) is $1/\sqrt{\pi}$ times the scale length of
    the exponential function (\widthkpc), the areas under the two
    curves are the
    same. }

    \label{fig:peak}
\end{figure}

The density of gas along the orbit of the cluster has a sharp rise at
the point where the shock occurs and an exponential-like decrease
after the shock (see our Fig.~\ref{fig:peak} and Fig.~5 of
\citealt{1969ApJ...158..123R}).  In reality, the density increase
  is bit smoother than shown in Fig.~\ref{fig:peak}. It is hard, if
  not impossible, to derive a suitable potential that describes a
  realistic density function. Therefore, we assume for simplicity a
  one-dimensional density perturbation of Gaussian form
\begin{equation}
\rho(Y)=\rho_0\,\exp(-Y^2/H^2),
\label{eq:rho}
\end{equation}
where $H$ is the scale length and $Y$ is the azimuthal distance. 
In Section~\ref{sec:theory} we show that for an impulsive perturbation
the compressive tidal forces on the cluster scale with the integral of
$\rho(Y)$ with respect to $Y$, where $Y$ is the direction of motion of
the cluster. A Gaussian function with a integral equal to that of an
exponential function has the same compressive effect on the
cluster. We emphasise that this is only true when for both
perturbations hold that they are impulsive. In practise, this implies
that the majority of the surface should be contained within the same
$Y$ extent for both functions. That is, a very broad gaussian with the
same surface as a very peaked gaussian will have a different effect on
the cluster, because the broad gaussian will heat the cluster
adiabatically, while the peaked gaussian will heat the cluster
impulsively. For the exponential function
$\rho(Y)=\rho_0\,\exp(-Y/\widthkpc)$, for $Y>0$, and the Gaussian
function of equation~(\ref{eq:rho}), the surfaces are equal when
$H=\widthkpc/\sqrt{\pi}$. With Table~\ref{tab:parameters} we thus find
that $H\simeq150\,$pc for the gaseous arms.

 By using Poisson's law and equation~(\ref{eq:rho}) we can derive the
related potential along the trajectory of the cluster

\begin{equation}
\phi(Y)=2\pi G\rho_0 H \left[\sqrt{\pi} Y \erf(Y/H) + H\exp(-Y^2/H^2)\right],
\label{eq:phi}
\end{equation}
where $G$ is the gravitational constant. The acceleration ($g(Y)$) that is
felt by the cluster due to spiral arm passage is equal to $-\dr
\phi/\dr Y$ and with equation~(\ref{eq:phi}) it follows that
\begin{equation}
g(Y)=-2\pi^{3/2} G\rho_0 H \erf(Y/H).
\label{eq:gy}
\end{equation}
The Gaussian density form (equation~\ref{eq:rho}) implies a constant
acceleration far from the centre of the density wave. This is not
physical for spiral arms, since the attractive force due to the arm
should decrease with distance from it and should reach zero in between
two arms. Therefore, we only consider the
effect of the density wave in the vicinity of its centre
($-2H<Y<+2H$), that is, where $g(Y)$ varies and hence tidal
forces are at work causing a compression of the cluster
(Section~\ref{sec:theory}).

The values of $\width$ and \widthkpc\ are always measured in the
tangential direction, that is, along the orbit of the cluster. The
width perpendicular to the spiral arm is therefore smaller and depends
on the pitch angle. Given the uncertainties in the arm width
measurements and since the typical radius of star clusters
($\sim3-4\,$pc) is much smaller than the $\widthkpc$ of the spiral
arm, it does not matter that clusters cross the arm with a certain
angle. In order to avoid having the pitch angle as an extra parameter,
we study perpendicular passages through a density wave. The effect of
the pitch angle is taken into account in the choice of the azimuthal
scale length $H$.

\section{Simple analytical estimates for one-dimensional tidal
  perturbations} 
\label{sec:theory}

\subsection{The impulsive approximation}
\label{subsec:impulsive}
The energy gain of a globular cluster crossing the Galactic disc was
derived by \citet{1972ApJ...176L..51O} using the impulse approximation.
They assumed that the stars do not move during the passage of the
density wave, that is, that \tcr\ is much longer than \Tcr.

Assume a one-dimensional acceleration along the $Y$-axis ($g(Y)$), a
cluster with its centre at $\Yc$ and an individual star at a distance
$y=Y-\Yc$ from the cluster centre.  The tidal acceleration of a star
due to the density wave is then
\begin{equation}
\frac{\dr \vy}{\dr t} = g(Y)-g(\Yc)\simeq y\frac{\dr g}{\dr Y}=\frac{y}{V}\frac{\dr g}{\dr t},
\label{eq:dvdt}
\end{equation}
where we have substituted $\dr Y = V \dr t$, with $V$ the relative
velocity between the cluster and the density wave and where $g$ is
expanded around the cluster centre. This is sufficiently accurate as
long as the cluster is much smaller than the width of the density
wave. Note that the tidal acceleration scales with the density
variation ($\rho(Y)$) since $g = -\dr \phi/\dr Y$ and
$\rho(Y)\propto\dr^2 \phi/\dr Y^2$ through Poisson's law. Therefore,
the tidal forces scale with $-\rho(Y)$ (equation~\ref{eq:dvdt}) and are
always directed inwards for a density increase. Combined with the
impulsive assumption,
\citet{1972ApJ...176L..51O} introduced the name {\it compressive
gravitational shock}. As mentioned in Section~\ref{sec:parameters},
the term shock was introduced due to the short duration of the
perturbation. This does not necessarily have to be true in our case, since for
cluster close to $\rcr$  \Tcr\ can be long.

Integrating equation~(\ref{eq:dvdt}) then yields an expression for the
velocity increment of a star of the form
\begin{equation}
\Delta \vy = \frac{y}{V}\left|\, g(Y_c)\,\right|_{-Y_{\rm m}}^{+Y_{\rm m}},
\label{eq:deltavy}
\end{equation}
where $Y_{\rm m}$ is the point where $g$ reaches its maximum $\gm$. If
$g$ is an odd function and does not change much for $Y<-Y_{\rm m}$ and
for $Y>+Y_{\rm m}$, the total energy gain per unit cluster mass after
a tidal perturbation is

\begin{equation}
\deimp = \frac{2\,g_{\rm m}^2\,\rrms}{3\,V^2},
\label{eq:deimp}
\end{equation}
where we have substituted $\yrms=\frac{1}{3}\rrms$ and \rrms\ is
the mean square position of stars in the cluster. From
equation~(\ref{eq:gy}) it follows that \gm\ is
\begin{equation}
\gm =2\pi^{3/2} G\rho_0 H.
\label{eq:gm}
\end{equation}
As mentioned in Section~\ref{subsec:potdens} the constant acceleration
at large distances from the Gaussian density is not physical in the case
of spiral arms. However, we are interested in the density change and
the related change in $g$ from $-\gm$ to $+\gm$ close to the centre of
the density wave is approximately correct.

\subsection{Validity of the impulsive approximation}

\subsubsection{Constant velocity assumption}
\label{subsec:velocity}
One of the assumptions made by \citet{1972ApJ...176L..51O} and later
in more thorough studies (see for example \citealt{1994AJ....108.1414W};
\citealt{1997MNRAS.291..717M}; \citealt{1997ApJ...474..223G,1999ApJ...513..626G}) on the tidal
perturbation due to the Galactic disc on globular clusters is that the
velocity, $V(Y)$, remains constant during the crossing. This is
probably not such a bad assumption, since globular clusters cross the
disc with high initial relative velocity ($\sim 200\,\kms$). For
spiral arm crossings, however, the relative velocity is between zero
at \rcr\ and almost $\pm\vdisc$ close to the galaxy centre and at
large distances from it. For low $|\vdrift|$ there is large relative
increase of velocity, making the assumption of a constant velocity
invalid (equation~\ref{eq:dvdt}) and
\ref{eq:deimp}). The solution to this can be obtained by 
expressing $V$ as a function of $g$ and integrating
equation~(\ref{eq:dvdt}). This, however, is even for this simple functional
form quite hard.

\begin{figure}
\centering
   \includegraphics[width=8.cm]{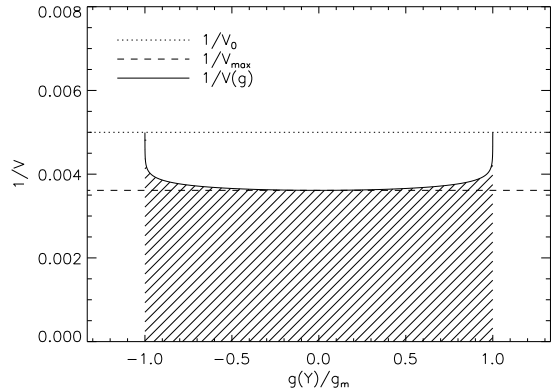}
   
    \caption{The variation of $1/V$ with $g(Y)$ (equation~\ref{eq:dvdt})
    for a constant velocity equal to $V_0$ and $\vmax$ (dotted and
    dashed, respectively) and for the true velocity $V(g)$ (full line),
    derived from the $\phi(Y)$ (equation~\ref{eq:phi}).}

    \label{fig:vmax}
\end{figure}

Alternatively, we can estimate the change in $V(Y)$ from $\phi(Y)$
(equation~\ref{eq:phi}). The variation of $V(Y)$ depends on $\phi(Y)$
as
\begin{equation}
V(Y) = \sqrt{V_0^2+2\left[\phi_0-\phi(Y)\right]},
\label{eq:vphi}
\end{equation}
wherex $\phi_0$ and $V_0$ are a reference potential and velocity,
respectively. The variation of $g(Y)$ follows from the spatial
derivative of $\phi(Y)$. We need to integrate $1/V$ with respect to
$g$ (equation~\ref{eq:dvdt}), to get the total $\Delta \vy$ of an
individual star. In Fig.~\ref{fig:vmax} we show the variation of $1/V$
with $g$ for a constant $V$ (dotted) line and one that considers an
acceleration due to the potential (full line). The shaded surface is
the result of the integration of $1/V$ with respect to $g$. This
surface is almost equal to $\simeq 2\gm/\vmax$.

Using equation~(\ref{eq:phi}), we can find an expression for $\Delta \phi$
for $Y_0>>H$ of the form
\begin{equation}
\Delta \phi=\phi(Y_0)-\phi(0)\simeq2\pi G\rho_0 H \left[\sqrt{\pi} Y_0 - H\right],
\label{eq:dphi}
\end{equation}
where $Y_0$ is the starting position of the cluster, where it has
velocity $V_0$ with respect to the density wave.
Equation~(\ref{eq:dphi}), combined with equation~(\ref{eq:vphi}), can be used
to find the value of $\vmax$.  This suggests that it is better to use
$\vmax$ instead of $V_0$ in equation~(\ref{eq:deimp}) to calculate the energy
gain.  In Section~\ref{sec7:nbody} we validate this with $N$-body
simulations.

\subsubsection{The effect of adiabatic damping}
\label{subsec:adiabatictheory}
 The value of \tcr\ depends on the distance of the stars to the
cluster centre ($r$). In the core of the cluster \tcr\ is much shorter
than close to the tidal radius (\rt). For stars with short period
orbits, the effect of the shock is largely damped due to adiabatic
invariances. Therefore, there is almost no net energy gain in the
core. On the other hand the stars in the outer region will largely be
heated impulsively. To define the transition between the
impulsive and adiabatic regions, we can define an adiabatic parameter
\citep{1987degc.book.....S}

\begin{equation}
x\equiv\frac{\omega H}{V},
\label{eq:x}
\end{equation}
where $\omega$ is the angular velocity of stars inside the
clusters. This is defined as $\omega(r)=\sigma(r)/r$, with $\sigma(r)$
the velocity dispersion of stars at position $r$. (Note that the
original definition of the adiabatic parameter by
\citealt{1987degc.book.....S} contains an additional factor 2. We
choose to define $x$ as the ratio of the time it takes the cluster to
cross a distance $H$ to the time it takes a star to cross a distance
$r$ inside the cluster, e.g. $\Tcr/\tcr$, as did
\citet{1999ApJ...513..626G}.) When $x<<1$ the term shock is justified,
while for $x>>1$ the perturbation is largely adiabatic.

\citet{1958ApJ...127...17S} gives an estimate of the conservation of
 adiabatic invariants in the harmonic potential approximation. This
 assumes all stars are initially at the same distance $r$ from the
 cluster centre and have the same oscillation frequency $\omega$. The
 energy gain for each star can then be written as a function of the
 result of the impulsive approximation (Section~\ref{subsec:impulsive})
 and an adiabatic correction factor $A(x)$ (equation~5 of
 \citealt{1995ApJ...438..702K})

\begin{equation}
\Delta E=\deimp\,A(x),
\label{eq:dea}
\end{equation}
where \deimp\ is given by equation~(\ref{eq:deimp}). The correction factor
$A(x)$ in the harmonic approximation is (for example
\citealt{1987degc.book.....S}, Eqs.~5-28)

\begin{equation}
\As=\exp(-2x^2).
\label{eq:aspitzer}
\end{equation}
We refer to \As\ as the Spitzer correction. (The original definition of
\As\ by Spitzer has a factor 0.5 instead of 2, because of the
different definition of $x$ in equation~(\ref{eq:x}).)

\citet{1994AJ....108.1398W,1994AJ....108.1403W,1994AJ....108.1414W}
showed that this exponential decrease of the energy with $x$
underestimates the heating effect.  This is due to the fact that the
basic assumption of the harmonic potential approximation is not valid
when the system has more than one degree of freedom, so that small
perturbations can still grow. When the cluster is represented as a
multidimensional system of nonlinear oscillators, some perturbation
frequencies become commensurable with the oscillator frequencies of
stars. This results in a correction factor $A(x)$ that is not
exponentially small for large $x$ but, instead, has a power-law
dependence on $x$.  The simplest form, as shown by
\citet{1997ApJ...474..223G}, can be written as

\begin{equation}
\Aw=(1+x^2)^{-3/2}.
\label{eq:aweinberg}
\end{equation}
We refer to \Aw\ as the Weinberg correction. 

Besides this shift in energy ($\Delta E$), there is also a quadratic
term that affects the dispersion of the energy spectrum of the
cluster. The effect of shock-induced relaxation was mentioned by
\citet{1973ApJ...183..565S} and later studied in more detail by
\citet{1995ApJ...438..702K}. It is beyond the scope of this study to
include all this detailed physics and we refer the reader to the
aforementioned papers for details.

In Section~\ref{sec7:nbody} we will compare results from $N$-body
simulations to both adiabatic correction factors
(Eqs.~\ref{eq:aspitzer} and \ref{eq:aweinberg}).

\section{$N$-body simulations}
\label{sec7:nbody}

In this Section, we confront the simple analytical estimates given in
Section~\ref{sec:theory} with a    
series of $N$-body simulations of clusters that cross a density wave
of Gaussian form (equation~\ref{eq:rho}). The potential and acceleration are derived from the
one-dimensional Gaussian density profile, as described in
Section~\ref{subsec:potdens}.

\subsection{Set-up of the simulations}

\subsubsection{Description of the code}
The $N$-body calculations were carried out by the \texttt{kira} integrator,
which is part of the \texttt{Starlab} software environment
(\citealt{1996ApJ...467..348M}; \citealt{2001MNRAS.321..199P}). {\texttt Kira}
uses a fourth-order Hermite scheme and includes special treatments of
close two-body and multiple encounters of arbitrary complexity. The
special purpose {\it GRAPE-6} systems \citep{2003PASJ...55.1163M} of the 
Observatoire de Marseille and of the University of Amsterdam were used to
accelerate the calculation of gravitational forces between stars. 

\subsubsection{Units and scaling}
\label{subsec:units}
The cluster energy per unit mass (\ec) is defined as $\ec=\eta
G\Mcl/(2\rh)$, with $\eta\simeq0.4$, \Mcl\ is the mass of the cluster and
\rh\ is its half-mass 
radius, depending on the cluster model.  All clusters are scaled to
$N$-body units, such that $G = \Mcl = 1$ and $\ec = -1/4$, following
\cite{1986LNP...267..233H}. The virial radius ($\rv$) is the unit of
length and follows from the scaling of the energy, since $\rv\equiv
G\Mcl/(2|U|)=1$, where $U$ is the potential energy per unit of cluster
mass.  We assume virial equilibrium at the start of the simulation,
which implies $U=2\ec$ and, therefore, $\rv(t=0)=1$.

\subsubsection{Cluster parameters}
\label{subsec:clusterpars}
For the density distribution of the cluster we assume a
\citet{1966AJ.....71...64K} profile, which fits the radial
luminosity profile of our Galactic clusters. Open clusters in our
Galaxy have a lower concentration index than globular
clusters. Defining the concentration index as $c = \log(r_t/r_c)$,
where $\rt$ is the tidal radius and $\rc$ is the core radius of the
cluster, we find that typical values for open clusters are in the
range  $0.5\,<\,c\,<\,0.9$
\citep{1962AJ.....67..471K} or even lower (\citealt{1987gady.book.....B}). A
concentration index $c=0.7$ corresponds to a dimensionless central
potential depth of $W_0=3$. The average concentration index of
globular clusters is $c\simeq1.5$ \citep{1996yCat.7195....0H},
corresponding to $W_0=7$. Here, we adopt a dimensionless central
potential depth of $W_0 = 5$. This corresponds to a concentration
index of $c=\log(\rt/\rc)\simeq1.03$. For a $W_0=5$ cluster it follows
that $\eta=0.41$. In $N$-body units, the corresponding radii are
$\rh=0.81$ and $\rrms=1.36$, respectively
(Section~\ref{subsec:units}).

\subsubsection{Code testing}
\label{subsec:testing}
\begin{figure}
\centering
   \includegraphics[width=8.cm]{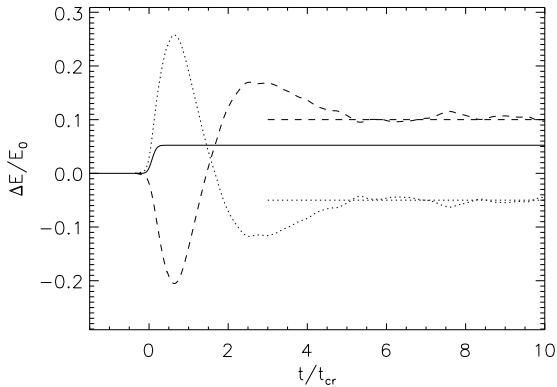}
   
    \caption{Variation of fractional energy gain ($\Delta E/|E_0|$), shown
    as a full line and of the kinetic and potential component shown as
    dotted and dashed lines, respectively. The cluster has $N=65\,536$
    particles and started in virial equilibrium. Parameters for the
    density wave are described in the text (Section~\ref{subsec:testing}).}

    \label{fig:de1}
\end{figure}

To test our code, we run a single perturbation with $H=100$ and
$V_0=200$ and $\rho_0$ chosen such, that $\de=0.05$
(Eqs.~\ref{eq:deimp} and \ref{eq:gm}). In this example, the cluster
consists of $N=65\,536$ equal mass particles. It is initially
positioned at $Y_0=-2H$ and its velocity is directed towards the
density wave. When the cluster is at $+2H$, there are almost no tidal
forces anymore, but there is still an acceleration towards the density
wave (equation~\ref{eq:gy}). To prevent a second crossing, we turn the
external tidal field off. The cluster is evolved for an additional
10\tcr\ after the perturbation. In Fig.~\ref{fig:de1} we show the
variation of the energy gain\footnote{When we discuss cluster energy,
we refer to the sum of the {\it internal} potential and {\it internal}
kinetic energy, that is, where the energy of the centre of mass motion
and the contribution of the external tidal field have been
subtracted.}  ($\Delta E$) as a full line and of the internal kinetic
energy ($\Delta T$) and potential energy ($\Delta U$) as dotted and
dashed lines, respectively. Due to the compressive nature of the tidal
forces, all stars gain kinetic energy in the $-y$ direction, resulting
in an increasing $T$. Consequently, the stars move deeper in the
potential of the cluster, resulting in a decreasing $U$. The cluster
is in virial equilibrium before the perturbation, that is, the virial
ratio is equal to one ($Q=-2T/U=1$). The value of $Q$ increases due to
the perturbation. After the perturbation the cluster revirialises, due
to relaxation, reducing $Q$ initially to $Q<1$, around
$t\simeq3\tcr$. After a few oscillations the virial ratio is close to
one again, leaving a cluster with $\Delta T=-\Delta E$ and $\Delta
U=+2\Delta E$.

\subsection{Parameters of the runs}
\label{subsec:parameters}
\begin{figure}
\centering
   \includegraphics[width=8.cm]{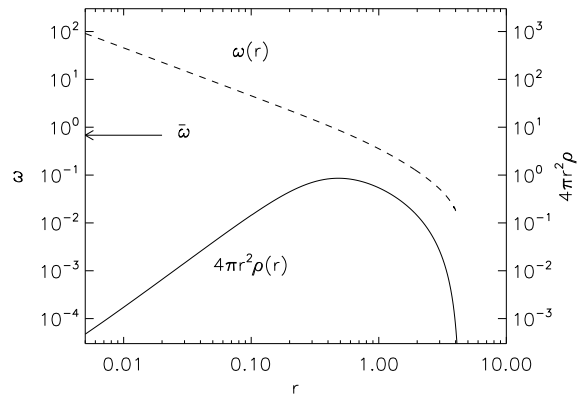}
   
    \caption{Frequency $\omega$ (full line, scale on the left) and the
    mass contribution $4\pi r^2\rho(r)$ as a function of $r$ (dashed
    line, scale on rigth). The value of equation~(\ref{eq:omegabar}) is
    indicated with an arrow.}

    \label{fig:omega}
\end{figure}

We choose to use the same clusters as before, but now with $N$=8k. The
value of $\omega$ varies strongly within the cluster, from
$\gtrsim100$ deep in the core down to $\sim0.01$ at $\rt$. We,
therefore, calculate the mass weighted mean value of $\omega$,
i.e. $\bar{\omega}$, by numerically solving

\begin{equation}
\bar{\omega}=\frac{4\pi}{\Mcl}\int_0^{\rt}\,r^2\rho(r)\omega(r)\dr r.
\label{eq:omegabar}
\end{equation}
For a King profile with $W_0=5$, $\bar{\omega}=0.68$. This is about a
factor of two higher than $1/\tcr$, with $\tcr=2\sqrt{2}$ at $\rv$ (in
$N$-body units). The radius at which $\omega(r)=0.68$ is slightly inside
the half-mass radius:
$r(\omega=\bar{\omega})=0.72\rh$. Fig.~\ref{fig:omega} illustrates the
variation of $\omega(r)$ (dashed line) and on $4\pi r^2\rho(r)$ (full
line) for a $W_0=5$ cluster.

The value of $\rho_0$ was chosen such that the predicted $\deimp$
(equation~\ref{eq:deimp}) is always 1/8 of $|\ec|$, which is $+1/32$
(in $N$-body units) (Section~\ref{subsec:units}). This is possible
since $\rho_0$ is a free parameter in Eqs.~\ref{eq:deimp} and
\ref{eq:gm} after $V_0$ and $H$ are defined by our grid.

\begin{figure*}
\centering
   \includegraphics[width=15.cm]{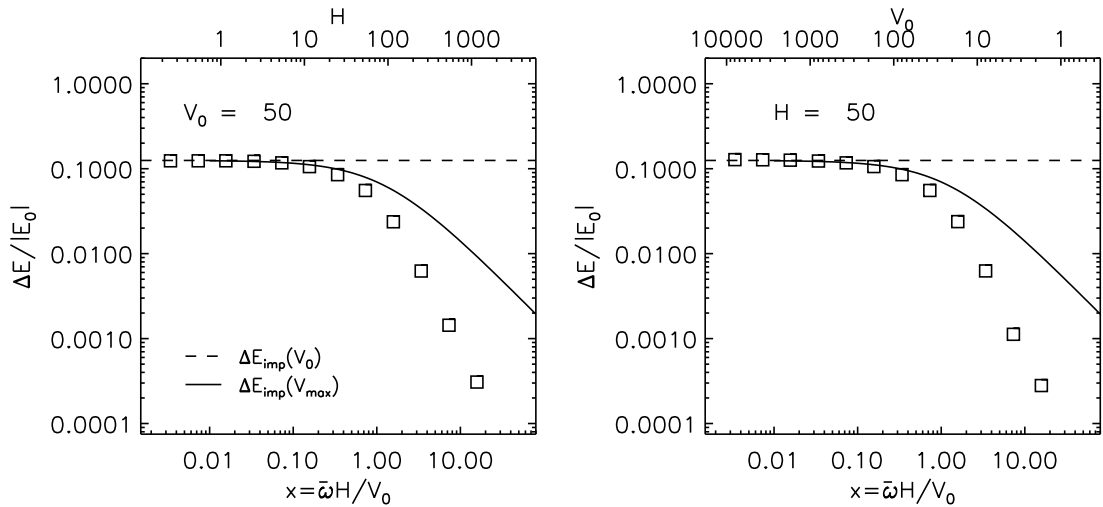}
   
    \caption{Fractional energy gain (\de) as a function of the
    adiabatic parameter $x$ (equation~\ref{eq:x}) resulting from the
    $N$-body simulations described in Section~\ref{subsec:parameters}
    (squares), the impulsive approximation (equation~\ref{eq:deimp}) based
    on $V_0$ (dashed line) and on $\vmax$ (full line). In the left
    panel the value for $V_0$ was fixed and $H$ was varied to vary
    $x$. In the right panel it was done the other way around. The two
    series give the result.}

    \label{fig:dehv}
\end{figure*}

The cluster is positioned initially at a distance $Y_0 = -2H$ with a
velocity $V_0$ in the $Y$-direction.

\subsection{Testing the validity of the impulsive approximation}
\label{subsec:velocitysim}

We ran two series of 12 simulations, both for the same sequence of $x$
values (equation~\ref{eq:x}), ranging from 0.003 to 15. This was achieved,
in the first series by fixing $V_0=50$ and varying $H$ and in the
second series by fixing $H=50$ and varying $V_0$ accordingly. Simple
analytical estimates relying on the impulsive approximation
(equation~\ref{eq:deimp}) give that the energy gain $\Delta E$ scales as
$1/V_0^2$ and is thus independent of $x$. If, however, we use in
equation~(\ref{eq:deimp}) $1/\vmaxsq$ instead, we find that $\Delta E$
decreases with increases $x$.

In Fig.~\ref{fig:dehv} we show $\de$ as a function of $x$, defined in
terms of $V_0$. The result for constant $V_0$ and $H$ are shown in the
left and right panel, respectively. The square symbols are the results
of the $N$-body simulations.  The dashed line shows the prediction of
$\deimp$ (equation~\ref{eq:deimp}) based on a constant velocity. Excellent
agreement between the simulations and the analytical predictions using
the impulsive approximation is found for $x\lesssim0.03$, for both
series. With the full line we show the predicted energy gain when we
use $\vmax$ in equation~(\ref{eq:deimp}), which we derived from
Eqs.~\ref{eq:vphi} and \ref{eq:dphi}. Indeed the energy gain decreases
for larger $x$, since $\vmax/V_0$ increases for perturbations with
longer duration. This prediction of $\deimp(\vmax)$ explains the
results of the simulations very nicely for $x\lesssim0.5$. For
$x\gtrsim0.5$ the energy gain in the simulations is lower than
predicted by the impulsive approximations. This is caused by adiabatic
invariances in the cluster core
(Section~\ref{subsec:adiabatictheory}). The two sequence give nearly
identical results, which suggests that it is the ratio $H/V_0$, or the
duration of the perturbation that controls the energy gain together
with $\rho_0$. In Section~\ref{subsec:adiabatic} we compare the
results of the simulations to the theoretical predictions that include
adiabatic damping.

\subsection{Adiabatic invariances}
\label{subsec:adiabatic}
\begin{figure*}
\centering
   \includegraphics[width=15cm]{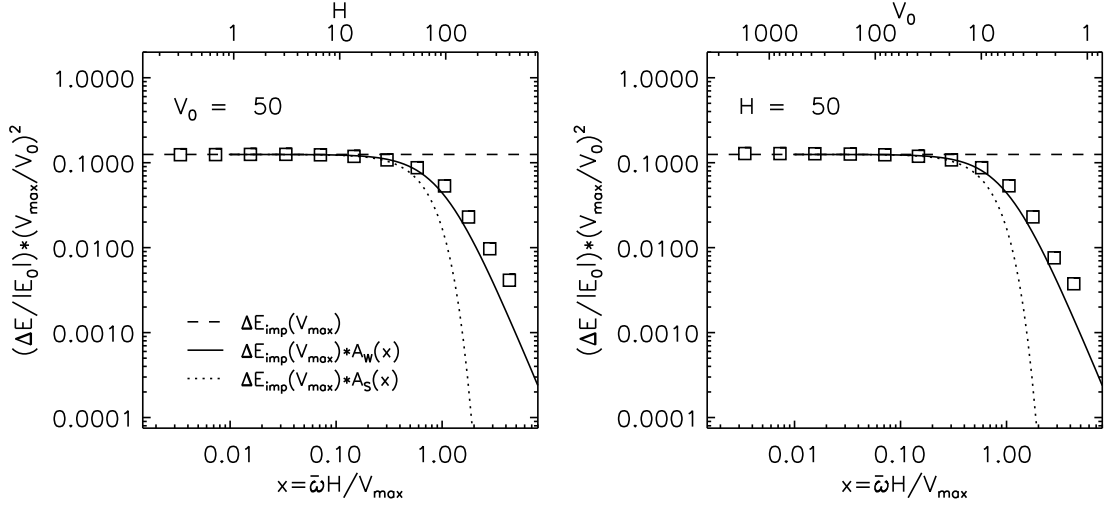}
   
    \caption{Same as in Fig.~\ref{fig:dehv}, but now $\de$ is
    multiplied by $(\vmax/V_0)^2$ to make the predicted energy gain using
    the impulsive approximation constant for all
    simulations. Deviations are now caused by adiabatic damping
    (Section~\ref{subsec:adiabatictheory}). The impulse approximation
    (dashed line) now predicts a constant \de\
    (equation~\ref{eq:deimp}). The adiabatic correction of Spitzer
    (equation~\ref{eq:aspitzer}) and Weinberg
    (equation~\ref{eq:aweinberg}) are shown as 
    a dotted line and as a full line, respectively.  }

    \label{fig:de_with_ad}
\end{figure*}

To quantify deviations from the impulsive approximation, that is, the
effect of damping by adiabatic invariances, we show in
Fig.~\ref{fig:de_with_ad} the resulting $\de$ of the simulations from
Section~\ref{subsec:velocitysim}, but now using
\vmax\ in the definition of $x$ (equation~\ref{eq:x}) and $\deimp$
(equation~\ref{eq:deimp}). The values of \de\ from the simulations are
multiplied by $(\vmax/V_0)^2$. If all simulations were in the impulse
regime, the energy gain would be constant ($\de=1/2$), which is
shown as a dashed line. (Note that the dashed line in
Fig.~\ref{fig:de_with_ad} is the same as the full line in
Fig.~\ref{fig:dehv}). The theoretical predictions with adiabatic
correction are shown for the result of Spitzer (equation~\ref{eq:aspitzer})
and Weinberg (equation~\ref{eq:aweinberg}) as dotted and full lines,
respectively. As was already clear from Fig.~\ref{fig:dehv}, the
simulations can be well explained by the impulsive approximation for
$x\lesssim0.5$. The decrease of \de\ for $x\gtrsim0.5$ is well explained by
the predictions made by
\citet{1994AJ....108.1398W,1994AJ....108.1403W,1994AJ....108.1414W},
that is, equation~(\ref{eq:aweinberg}) (full line).

From Table~\ref{tab:parameters} we find that the values of $x$
for a cluster with $\bar{\omega}=1\,$Myr and stellar arms at
$R=[1,3,5]\,$kpc in a grand design spiral are $\sim[3,10,50]$,
respectively. Combined with the results of Figs.~\ref{fig:dehv} and
\ref{fig:de_with_ad} this support our assumption that the stellar arms do
not contribute to the disruption of clusters
(Section~\ref{sec:parameters}). 

For the gaseous arms close to \rcr\ the values of $|\vdrift|$ are low,
and $x\gtrsim1$ (Section~\ref{sec:parameters}). It is, therefore,
important to include a correct description of adiabatic damping to
understand the effect of spiral arm perturbations on
clusters. Including the simple correction as derived by Spitzer
(equation~\ref{eq:aspitzer}) underestimates the energy gain of
perturbations with low $|\vdrift|$. The impulsive approximation
(equation~\ref{eq:deimp}) overestimates the energy gain by orders of
magnitude.

\section{Cluster mass loss and disruption by spiral arm perturbations}
\label{sec7:disruption}

\subsection{Energy gain {\it vs.} mass loss}
\label{subsec:dedm7}
The energy gained in a compressive perturbation is absorbed mainly by
the stars in the outer parts, since for these stars $r$ is large
(Eqs.~\ref{eq:dvdt}, \ref{eq:deltavy}). This implies that the
fractional energy gain (\de) is not necessarily equal to the
fractional mass loss\footnote{We define the mass loss, $\Delta
M$, to be positive. {\it Bound} is defined as having a velocity lower
than the escape velocity, which is based on the cluster potential
only, that is, corrected for the presence of the external potential.} (\dm).
In fact,
\de\ is higher than \dm, since stars escape with velocities higher
than the escape velocity. Since this energy is taken by escaping
stars, it can not be used to unbind more stars. This was also found
for encounters between star clusters and GMCs
\citep{2006MNRAS.tmp..808G}.

\begin{figure}
\centering
    \includegraphics[width=8cm]{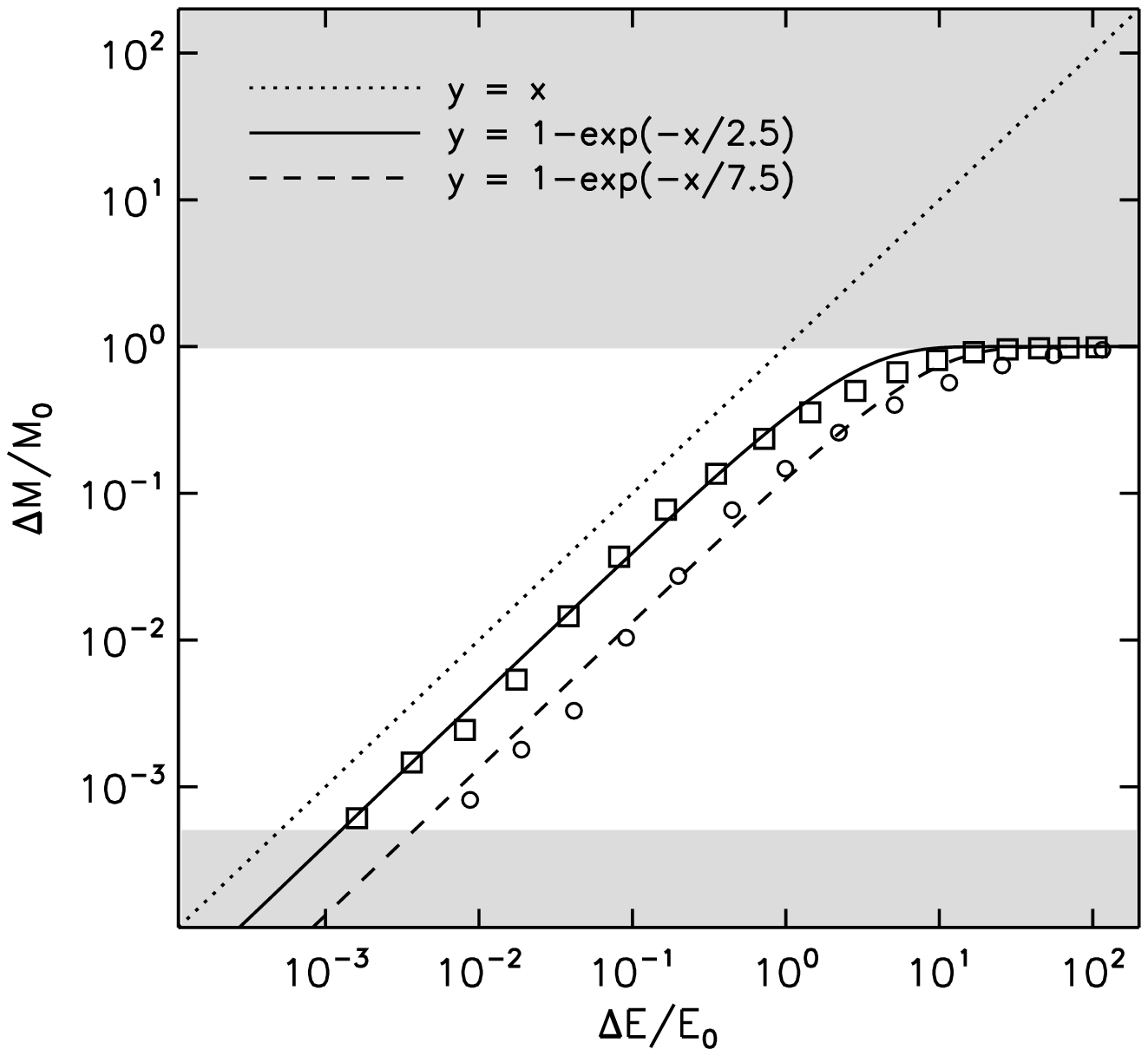}

    \caption{Fractional mass loss defined as the fractional loss of
    bound stars (\dm) {\it vs.} fractional energy gain (\de) from the
    simulations (circles). The dashed line is a simple functional form
    that makes \dm\ a factor of 7.5 lower than \de\ for $\de<1$ and
    flattens to 1 for $\de>1$.  The \dm\ defined as the number of
    stars beyond \rt\ is shown as squares. A functional fit is shown
    as a full line. A one-to-one relation is shown as dotted line. The
    results of the simulations are expected to be between the grey
    shaded regions, since the maximum value for \dm\ is 1,
    corresponding to the total disruption of the cluster and the
    minimum value is 1/8192, when a single star got unbound by the
    perturbation.  }

    \label{fig:dedm}
    
\end{figure}

We simulate perturbations with \de\ between $10^{-3}$ and 100 for
clusters with $N=2048$. The values of $H$ and $V_0$ were fixed at 10
and 100, respectively. The value of $\rho_0$ was varied to achieve the
desired energy gain. The simulations are continued without a tidal field for
five crossing times after the perturbations. The final
number of unbound stars is then compared to
\de\ from the simulations. In Fig.~\ref{fig:dedm} we show \dm\ as a
function of \de\ following from the simulations (circles). The dotted
line shows a one-to-one relation. For $\de<1$ the relation between
\dm\ and \de\ is almost linear, but with \dm\ a factor of 7.5 lower
than \de. For $\de>1$ the relation flattens to $\dm=1$, that is, to a
value corresponding to a 
completely unbound cluster. This relation can be described well by
a simple function of the form 
\begin{equation}
\dm=1-\exp(-f\de),
\label{eq:simple}
\end{equation}
 with $f\simeq1/7.5$ (dashed line). The reason for the deviation
of a one-to-one relation at $\de\simeq1$ is that the assumption of the
impulsive approximation is not valid anymore for such strong
perturbations. The stars in the centre of the cluster are heated
adiabatically (Fig.~\ref{fig:omega}) and remain bound after the
perturbation. A value of \de\ of almost $\sim$10 is necessary to completely
unbind the cluster with a single passage.

Stars that do not get unbound can get in higher energy orbits, where
in the presence of a tidal field they can be beyond \rt. In these
simulations we ignore the Galactic tidal field, but we can follow the
number of stars that is pushed over \rt. For this, we adopt a
physically more relevant definition of \dm\ that considers all unbound
stars as well as stars that are still bound to the cluster, but have
positions larger than
\rt. The relation between this \dm\ and \de\ is shown as squares in
Fig.~\ref{fig:dedm}. These values are higher, but still a factor of
2.5 lower than \de, and the results can be approximated by
equation~(\ref{eq:simple}), with $f\simeq1/2.5$ (full line). To
translate the energy gains predicted from the analytical estimates to
mass loss of the cluster we adopt equation~(\ref{eq:simple}) with
$f=0.4$. With this we estimate the mass loss and resulting
disruption time of clusters due to spiral arm perturbations in
Section~\ref{subsec7:disruption1}.

\subsection{The cluster disruption time}
\label{subsec7:disruption1}

\subsubsection{A toy model for spiral arms in the disc}
\label{subsec:toymodel}
The strength of a spiral arm perturbation on a cluster depends on
\vdrift, $H$ and $\rho_0$. In Sections~\ref{subsec:gaseous} and
\ref{subsec:potdens} and Table~\ref{tab:parameters} we adopted a
constant arm-interarm density contrast $\Gamma=10$ and scale length
 of $H=150\,$pc for gaseous arms, leaving
\vdrift\ as main $R$ dependent variable. 

From equation~(\ref{eq:rho0}) and $\meanrhoh$ from
Section~\ref{subsec:gaseous} we find $\rho_0=0.33\,\msunpccube$. The
value of \vmax\ in the arm is found from \vdrift\ and by considering an
acceleration by the arm from $Y_0=-2H$ to $Y=0$
(equation~\ref{eq:vphi}). With these parameters and
Eqs.~\ref{eq:dea} and \ref{eq:aweinberg} we can now calculate
the energy gain for clusters that cross spiral arms at different $R$.

With the relation between \dm\ and \de\ from
Section~\ref{subsec:dedm7}, that is, equation~(\ref{eq:simple}) with the
factor $f=0.4$, we can estimate the mass loss of the cluster at
different $R$. We assume here that \dm\ is independent of $N$, which
can be justified  since the derivations of \de\
(equations.~\ref{eq:deimp} and \ref{eq:dea}) are independent of $N$ as
well. They do, however, depend on the distribution of stars in the clusters,
that is, on the cluster density profile or the concentration index. There
might be a small $N$ dependence in the boundary between the impulsive
and the adiabatic regime, since clusters with low $N$ are able to
adjust faster to a changing potential.

\subsubsection{The cluster disruption time}
\label{subsec7:disruption}

\citet{1972ApJ...176L..51O} define the cluster disruption time 
due to many repeated disc shocks ($t_{\rm dd}$) using the impulsive
approximation by dividing the initial cluster energy $\ec$ by the
energy injection per unit time: $t_{\rm dd}=|\ec|/\langle\dr E/\dr
t\rangle$. For the cluster energy they assume $\ec=-0.2\,G\,\Mcl/\rh$
(Note that this implies $\eta=0.4$, see
Section~\ref{subsec:units}). Combined with equation~(\ref{eq:deimp}) and the
fact that clusters have two disc passages per orbital period ($P$),
\citet{1972ApJ...176L..51O} express $t_{\rm dd}$ as

\begin{equation}
t_{\rm dd} = \frac{3\,G\,\Mc\,P\,V^2}{20\gmsq\,\rhcube},
\label{eq:tshock1}
\end{equation}
This time scale is defined as the time needed to unbind the cluster by
(periodically) injecting energy in the cluster by disc shocks. Note
that here we have assumed $\rhsq=\rrms$. This is true only for very
concentrated clusters. 

This expression can be rewritten to derive a disruption time due to
periodic spiral arm perturbations ($\tsh$). 
We explicitly use the mass loss per spiral arm passage
(equation~\ref{eq:simple}), which we approximate as $\dm=f(\de)$,
with $f\simeq0.4$ (Section~\ref{subsec:dedm7}). The expression for \tsh\ can then be derived from

\begin{equation}
\tsh = \frac{\Delta t}{f}\frac{|\ec|}{\Delta E},
\label{eq:tsh}
\end{equation}
with $\Delta E$ from equation~(\ref{eq:dea}) (with the adiabatic
correction factor) and $\Delta t=\tdrift(R)$. Using the definition of
\ec\ from Section~\ref{subsec:units}, we can write \de\ as 

\begin{eqnarray}
\de & = & \frac{2\gmsq\rrms}{3\vmaxsq}\frac{2\rh}{\eta G\Mcl}\Aw \nonumber\\ 
    & = & \frac{4\gmsq}{3\vmaxsq}\frac{\rhcube}{\eta
    G\Mcl}\frac{\rrms}{\rhsq}\Aw,
\label{eq:de7}
\end{eqnarray}
where we included a factor $\rrms/\rhsq$ to get a dependence on
$\Mcl/\rhcube$ as in equation~(\ref{eq:tshock1}). For a cluster with
$W_0=5$  $\rrms/\rhsq\simeq2$.

\begin{figure*}
\centering
   \includegraphics[width=18cm]{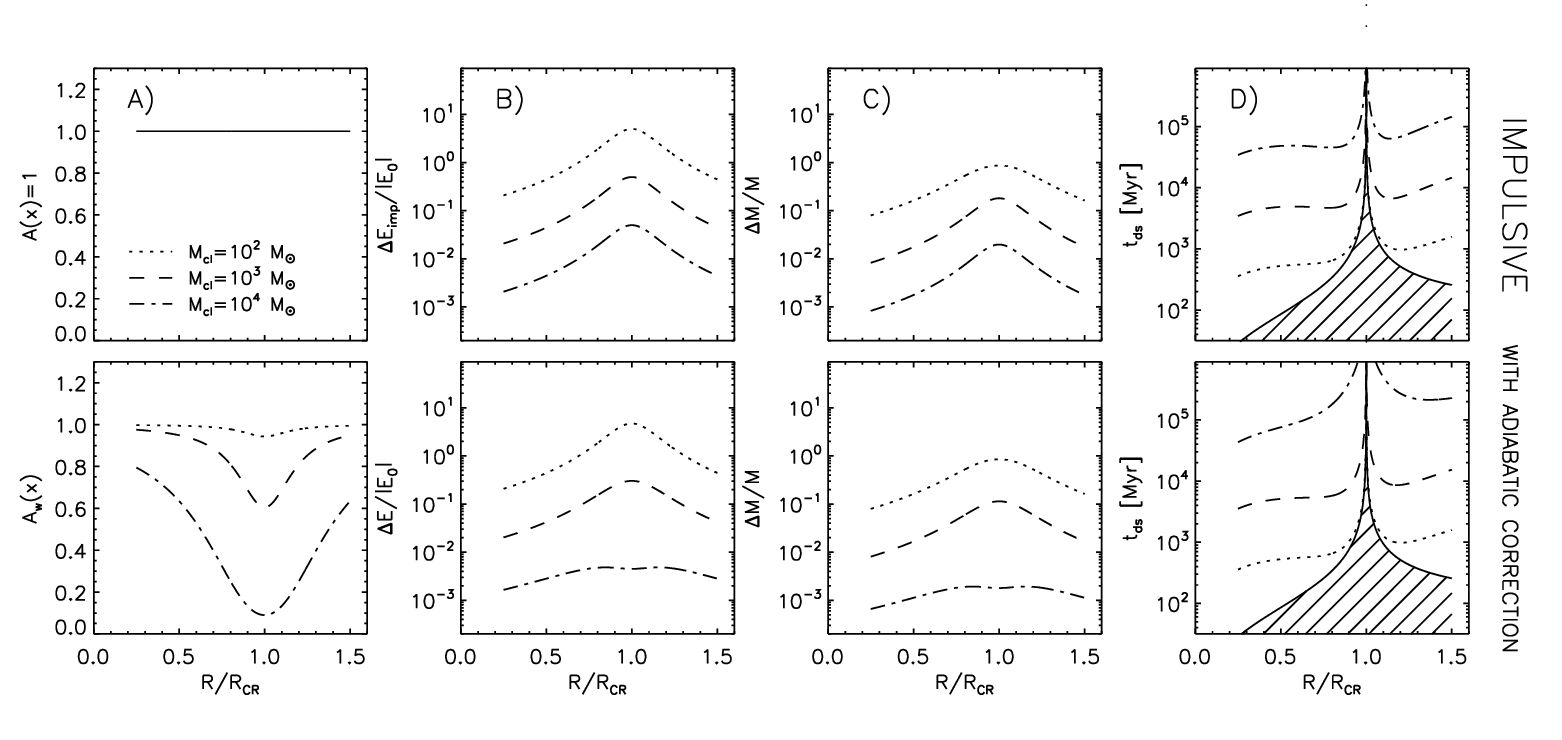}
   
    \caption{ Evolution of the cluster parameters with $R$ in a
    two armed grand design spiral galaxy. In the top row results are
    shown using the impulsive approximation, that is $A(x)=1$, and in
    the bottom panel we used \Aw\ of
    equation~(\ref{eq:aweinberg}). The adiabatic correction factor is
    shown in panel A. Panel B shows \de\ for a single shock at
    different $R$. The peak around \rcr\ is because of a minimum in
    \vmax\ around \rcr. From \de\ we derive \dm\ through
    equation~(\ref{eq:simple}) and the result is shown in panel C. The
    disruption time at different $R$, $\tsh(R)$
    (equation~\ref{eq:tshock}), is shown in panel D. The results for
    clusters with masses of $10^2\,\msun\,10^3\,\msun$ and
    $10^4\,\msun$ are shown with dotted, dashed and dotted-dashed
    lines, respectively. The hashed region indicated the region
    where \tsh\ is smaller than \tdrift, and therefore \tsh\ from
    equation~(\ref{eq:tshock}) can not exist.} 

    \label{fig:panel1}
\end{figure*}

The  adiabatic  correction  factor (\Aw) of equation~(\ref{eq:de7})  is
calculated using  the results  of Section~\ref{sec7:nbody}. We  use the
parameters  for  the  $W_0=5$   cluster  as  described  in
Sections~\ref{subsec:clusterpars} and
\ref{subsec:parameters}, that is, $\eta=0.41$ and $\bar{\omega}=0.58$. We consider
clusters with $\Mcl=10^2\,\msun,\,10^3\,\msun$ and $10^4\,\msun$ all
with $\rh=3.75\,$pc. The constant radius is based on recent
observations of young clusters in spiral galaxies
\citep{2004A&A...416..537L}. Using the scaling relations of
Section~\ref{subsec:units}, the values of $\bar{\omega}$ for these
clusters are $0.046,0.14$ and $0.46\,$Myr$^{-1}$, respectively. In
panel A of Fig.~\ref{fig:panel1} we show the result for the impulsive
approximation, that is $A(x)=1$, in the top panel and Weinbergs
correction factor ($\Aw$) in the bottom panel.

In panel B of Fig.~\ref{fig:panel1} we show the energy gain of
equation~(\ref{eq:de7}) using values of $A(x)$ from panel A. For the
cluster with $\Mcl=10^2\,\msun$ inclusion of the correct \Aw\ does not
make much difference. For the cluster with $\Mcl=10^3\,\msun$, \Aw\
decreases to 0.2 at \rcr\ and for the most massive cluster
($\Mcl=10^4\,\msun$) the effect of the spiral arm passage is largely
damped for almost all values of $R$. This is because  \tcr\ is
shorter in a more massive cluster, assuming a constant radius.

If we translate \de\ to \dm\ with equation~(\ref{eq:simple}) we prevent
that the mass loss becomes larger than the cluster mass. Panel C
shows \dm\ in the impulsive approximation (top) and with the adiabatic
correction factor (bottom). The cluster with $\Mcl=10^2\,\msun$ can
get completely unbound by a single spiral arm passage (\dm=1).

An expression for \tsh\ can be obtained by combining
Eqs.~\ref{eq:tsh} and \ref{eq:de7}

\begin{equation}
\tsh(R) = \frac{3\vmaxsq}{4g_{\rm m}^2}\,\frac{\eta}{f}\frac{1}{\Aw}\frac{G\Mcl}{\rhcube}\frac{r_h^2}{\rrms}\,\tdrift,
\label{eq:tshock}
\end{equation}
where $\vmax\simeq|\vdrift(R)|$ (equation~\ref{eq:vdrift}) and $\tdrift=
2\pi R/(m|\vdrift|)$, with $m$ the number of spiral arms in the
galaxy. In panel D of Fig.~\ref{fig:panel1} we show the result of
\tsh\ as a function of $R$ for a two-armed spiral galaxy
($m=2$). For clusters inside the corotation radius, i.e. $R<\rcr$, the
value of \tsh\ is nearly constant. This is because we have assumed the
arm parameters not to vary with $R$
(Section~\ref{subsec:gaseous}). The variables that are a function of
$R$ in equation~(\ref{eq:tshock}) are \vmax\ and \tdrift. The product
$V^2_{\rm max}\tdrift$ from equation~(\ref{eq:tshock}) varies with $R$
approximately as $R|1-R/\rcr|$, since $\tdrift\propto R/|\vdrift|$ and
$\vmax\simeq|\vdrift|$ (equation~\ref{eq:vdrift}).  The product
$R|1-R/\rcr|$ increases from 0 at $R=0$ to a maximum at $\rcr/2$ to
decrease again beyond $R=\rcr/2$. Since at \rcr\ there is never a
spiral arm crossing \tsh\ is infinite there.

\section{Discussion and conclusions}
\label{sec:conclusion}

Much work has been done on the combined effect of stellar evolution, a
realistic stellar mass function and a spherically symmetric Galactic
tidal field to understand the dissolution of star clusters. The
resulting cluster disruption times of these models (for example
\citealt{1990ApJ...351..121C, 2000ApJ...535..759T, 2003MNRAS.340..227B}) are longer than the observed values for
open clusters in the solar neighbourhood \citep{2005A&A...441..117L}
and in the central region of M51 \citep{2005A&A...441..949G}. Our
study is aimed at understanding the effect of perturbation by spiral
arm passages and to see whether they can contribute to this
difference.

From the age distribution of open clusters in the solar neighbourhood
clusters, \citet{2005A&A...441..117L} derived a disruption time of
$1.3\times10^3\,$Myr for a $10^4\,\msun$ cluster, which they refer to
as $t_4$. A similar analysis was performed by
\citet{2005A&A...441..949G} using the age and mass distribution to
derive the disruption time of clusters in the central region of M51. They
find $t_4\simeq150\pm50\,$Myr for clusters with $R<6\,$kpc, that is,
within \rcr\ (Section~\ref{sec:parameters}). These observed disruption time scales can be compared to the
results of
\citet{2003MNRAS.340..227B}. They derive the disruption time for
clusters on a circular motion in the Galaxy and consider a realistic
stellar mass function and stellar evolution and the tidal field of the
Galaxy. They used a smooth analytical description of the Galaxy
potential and find $t_4 =6.9\times10^3\,$Myr for clusters in the solar
neighbourhood, that is, almost a factor of five longer than
what is observed. For the parameters of M51, the predicted value of $t_4$ is
ten times longer than the observed value.

\begin{figure}
   \includegraphics[width=8cm]{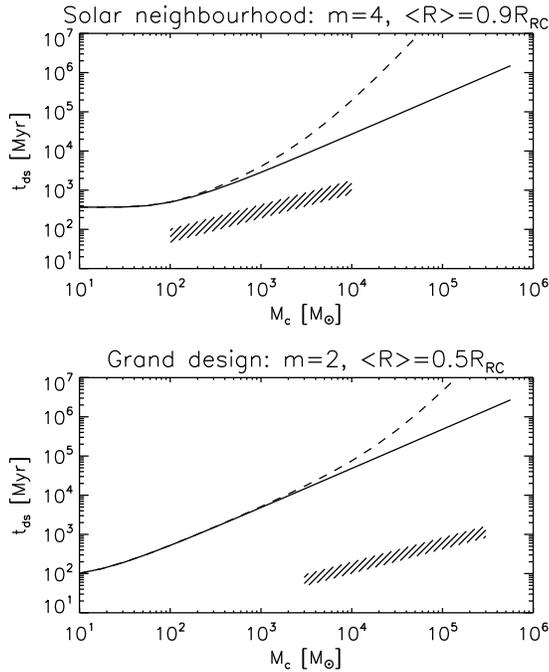}
       \caption{Disruption time due to spiral arm passages (\tsh) for
        different \Mcl\ in the solar neighbourhood (top panel) and in a
    gra	nd design spiral galaxy (bottom). The full lines show the
    result using the impulsive approximation, that is,
    equation~(\ref{eq:tshock}) with $A(x)=1$. The dashed lines show
    \tsh\ with $\Aw$ (Fig.~\ref{fig:panel1}). The shaded areas are the
    observed disruption times and mass range of the observed clusters
    from \citet{2005A&A...441..117L} (top) and
    \citet{2005A&A...441..949G} (bottom).}

    \label{fig:tdis}
\end{figure}

Neither the spiral arm perturbations on a simple cluster, without mass
function and stellar evolution (Section~\ref{subsec7:disruption}), nor
the effect of the Galactic tidal field on a realistic cluster can, on
their own, explain the observed disruption time. We can compare the
disruptive effect of the spiral arm perturbations to the observed
values. In Fig.~\ref{fig:tdis} we show the \tsh\ as a function of
cluster mass \Mcl\ for the solar neighbourhood (top panel) and for a
grand design spiral galaxy, such as M51 (bottom panel). The parameters
of Sections~\ref{sec:parameters} and \ref{subsec7:disruption} are
used, that is, the number of spiral arms is four ($m=4$) and
$R\simeq0.9\rcr$, with $\rcr$ the corotation radius, for the solar
neighbourhood and $m=2$ and $R\simeq0.5\rcr$ for the clusters in
M51. The full lines represent the resulting disruption times using the
impulsive approximation and the dashed lines show the result for \tsh\
when the correct adiabatic correction is applied. As we showed in
Section~\ref{subsec7:disruption}, the impulsive result applies to
clusters with $\Mcl \lesssim3\times10^3\,\msun$ in the solar
neighbourhood, because of the long crossing time \tcr\ of stars in
such clusters. For the cluster in a grand design spiral at $R=0.5\rcr$
the adiabatic correction is important for clusters of slightly higher
masses, since at that location \vdrift\ is higher
(equation~\ref{eq:vdrift}). In the solar neighbourhood \tsh\ becomes
constant for $\Mcl\lesssim10^2\,\msun$. This is because a single arm
crossing can destroy clusters with such low masses and \tsh\ becomes
equal to \tdrift. The disruption times from the observations of
\citet{2005A&A...441..117L} and
\citet{2005A&A...441..949G} for the solar
neighbourhood and M51 are indicated as hashed regions in the top and
bottom panel, respectively. The area of the hashed region represents
the errors in the observations and the mass range of the observed
cluster sample. Note that both studies found an empirical  mass
dependence of $\tdis\propto\Mcl^{0.6}$.

For the solar neighbourhood the disruption time due to the spiral arm
perturbations is about an order of magnitude higher than the observed
result.  For M51 the difference is almost three orders of
magnitude. This suggest that spiral arm perturbations contribute
little to the disruption of the (low mass) open clusters in the solar
neighbourhood and can definitely not explain the observed short
disruption time of clusters in M51 \citep{2005A&A...441..949G}. 
In Section~\ref{sec:parameters} we argued that the contribution of the
stellar arms, which we have ignored, is even less than that of the
gaseous component. This because the time it takes the cluster to cross
of stellar arms is much longer than the crossing time of stars in the
cluster. This causes the effect of heating by the stellar arms to be
largely damped adiabatically (Section~\ref{sec:theory}).

There should thus be a further disruptive agent, which we have not
considered so far. The central region of M51 has a rich population of
giant molecular clouds (GMCs) (for example
\citealt{1993A&A...274..123G}) whose velocity with respect to the
cluster is  relatively small ($\sim10\,\kms$). Even  at the relatively
large  distance from the  Galactic centre  of the  solar neighbourhood
there   are  GMCs  with   masses  $\gtrsim10^6\,\msun$   (for  example
\citealt{1987ApJ...319..730S}).


\section*{Acknowledgements}
We thank Oleg Gnedin, Nate Bastian, Henny Lamers, Douglas Heggie, Piet
Hut, Steve McMillan, Jean-Charles Lambert and Albert Bosma for useful
discussion and the referee, Martin Weinberg, for many useful
suggestions. This work was done with financial support of a {\it
Marie-Curie Training Fellowship} with number HPMT-CT-2001-00338, the
Royal Netherlands Academy of Arts and Sciences (KNAW), the Dutch
Research School for Astrophysics (NOVA, grant 10.10.1.11 to HJGLM
Lamers). EA and MG also acknowledge INSU/CNRS, the PNG and the OAMP
for funds to develop the computing facilities used for the simulations
in this paper. Part of the simulations were done on the MoDeStA
platform in Amsterdam. MG thanks the Sterrekundig Instituut ``Anton
Pannenkoek'' of the University of Amsterdam and the Observatoire de
Marseille for their hospitality during many pleasant visits.

\bibliographystyle{mn2e}

\begin{thebibliography}{}

\bibitem[\protect\citeauthoryear{{Athanassoula}}{{Athanassoula}}{1984}]{1984Ph%
R...114..321A}
{Athanassoula} E.,  1984, \physrep, 114, 321

\bibitem[\protect\citeauthoryear{{Athanassoula}}{{Athanassoula}}{1992}]{1992MN%
RAS.259..345A}
{Athanassoula} E.,  1992, \mnras, 259, 345

\bibitem[\protect\citeauthoryear{{Bastian}, {Gieles}, {Lamers}, {Scheepmaker}
  \& {de Grijs}}{{Bastian} et~al.}{2005}]{2005A&A...431..905B}
{Bastian} N.,  {Gieles} M.,  {Lamers} H.~J.~G.~L.~M.,  {Scheepmaker} R.~A.,
  {de Grijs} R.,  2005, \aap, 431, 905

\bibitem[\protect\citeauthoryear{{Baumgardt} \& {Makino}}{{Baumgardt} \&
  {Makino}}{2003}]{2003MNRAS.340..227B}
{Baumgardt} H.,  {Makino} J.,  2003, \mnras, 340, 227

\bibitem[\protect\citeauthoryear{{Binney} \& {Tremaine}}{{Binney} \&
  {Tremaine}}{1987}]{1987gady.book.....B}
{Binney} J.,  {Tremaine} S.,  1987, {Galactic dynamics}.
Princeton, NJ, Princeton University Press, 1987

\bibitem[\protect\citeauthoryear{{Bissantz}, {Englmaier} \&
  {Gerhard}}{{Bissantz} et~al.}{2003}]{2003MNRAS.340..949B}
{Bissantz} N.,  {Englmaier} P.,    {Gerhard} O.,  2003, \mnras, 340, 949

\bibitem[\protect\citeauthoryear{{Cayatte}, {Kotanyi}, {Balkowski} \& {van
  Gorkom}}{{Cayatte} et~al.}{1994}]{1994AJ....107.1003C}
{Cayatte} V.,  {Kotanyi} C.,  {Balkowski} C.,    {van Gorkom} J.~H.,  1994,
  \aj, 107, 1003

\bibitem[\protect\citeauthoryear{{Chernoff} \& {Weinberg}}{{Chernoff} \&
  {Weinberg}}{1990}]{1990ApJ...351..121C}
{Chernoff} D.~F.,  {Weinberg} M.~D.,  1990, \apj, 351, 121

\bibitem[\protect\citeauthoryear{{de Grijs} \& {van der Kruit}}{{de Grijs} \&
  {van der Kruit}}{1996}]{1996A&AS..117...19D}
{de Grijs} R.,  {van der Kruit} P.~C.,  1996, \aaps, 117, 19

\bibitem[\protect\citeauthoryear{{Dehnen}}{{Dehnen}}{2000}]{2000AJ....119..800%
D}
{Dehnen} W.,  2000, \aj, 119, 800

\bibitem[\protect\citeauthoryear{{Dehnen}}{{Dehnen}}{2002}]{2002ASPC..275..105%
D}
{Dehnen} W.,  2002, in {Athanassoula} E.,  {Bosma} A.,   {Mujica} R.,  eds, ASP
  Conf. Ser. 275: Disks of Galaxies: Kinematics, Dynamics and Peturbations {Our
  Galaxy}.
pp 105--116

\bibitem[\protect\citeauthoryear{{Drimmel}}{{Drimmel}}{2000}]{2000A&A...358L..%
13D}
{Drimmel} R.,  2000, \aap, 358, L13

\bibitem[\protect\citeauthoryear{{Drimmel} \& {Spergel}}{{Drimmel} \&
  {Spergel}}{2001}]{2001ApJ...556..181D}
{Drimmel} R.,  {Spergel} D.~N.,  2001, \apj, 556, 181

\bibitem[\protect\citeauthoryear{{Egusa}, {Sofue} \& {Nakanishi}}{{Egusa}
  et~al.}{2004}]{2004PASJ...56L..45E}
{Egusa} F.,  {Sofue} Y.,    {Nakanishi} H.,  2004, \pasj, 56, L45

\bibitem[\protect\citeauthoryear{{Fux}}{{Fux}}{2001}]{2001A&A...373..511F}
{Fux} R.,  2001, \aap, 373, 511

\bibitem[\protect\citeauthoryear{{Garcia-Burillo}, {Guelin} \&
  {Cernicharo}}{{Garcia-Burillo} et~al.}{1993}]{1993A&A...274..123G}
{Garcia-Burillo} S.,  {Guelin} M.,    {Cernicharo} J.,  1993, \aap, 274, 123

\bibitem[\protect\citeauthoryear{{Georgelin} \& {Georgelin}}{{Georgelin} \&
  {Georgelin}}{1976}]{1976A&A....49...57G}
{Georgelin} Y.~M.,  {Georgelin} Y.~P.,  1976, \aap, 49, 57

\bibitem[\protect\citeauthoryear{{Gieles}, {Bastian}, {Lamers} \&
  {Mout}}{{Gieles} et~al.}{2005}]{2005A&A...441..949G}
{Gieles} M.,  {Bastian} N.,  {Lamers} H.~J.~G.~L.~M.,    {Mout} J.~N.,  2005,
  \aap, 441, 949

\bibitem[\protect\citeauthoryear{{Gieles}, {Portegies Zwart}, {Baumgardt},
  {Athanassoula}, {Lamers}, {Sipior} \& {Leenaarts}}{{Gieles}
  et~al.}{2006}]{2006MNRAS.tmp..808G}
{Gieles} M.,  {Portegies Zwart} S.~F.,  {Baumgardt} H.,  {Athanassoula} E.,
  {Lamers} H.~J.~G.~L.~M.,  {Sipior} M.,    {Leenaarts} J.,  2006, \mnras, 371,
  793

\bibitem[\protect\citeauthoryear{{Gnedin}, {Lee} \& {Ostriker}}{{Gnedin}
  et~al.}{1999}]{1999ApJ...522..935G}
{Gnedin} O.~Y.,  {Lee} H.~M.,    {Ostriker} J.~P.,  1999, \apj, 522, 935

\bibitem[\protect\citeauthoryear{{Gnedin} \& {Ostriker}}{{Gnedin} \&
  {Ostriker}}{1997}]{1997ApJ...474..223G}
{Gnedin} O.~Y.,  {Ostriker} J.~P.,  1997, \apj, 474, 223

\bibitem[\protect\citeauthoryear{{Gnedin} \& {Ostriker}}{{Gnedin} \&
  {Ostriker}}{1999}]{1999ApJ...513..626G}
{Gnedin} O.~Y.,  {Ostriker} J.~P.,  1999, \apj, 513, 626

\bibitem[\protect\citeauthoryear{{Gonzalez} \& {Graham}}{{Gonzalez} \&
  {Graham}}{1996}]{1996ApJ...460..651G}
{Gonzalez} R.~A.,  {Graham} J.~R.,  1996, \apj, 460, 651

\bibitem[\protect\citeauthoryear{{Harris}}{{Harris}}{1996}]{1996yCat.7195....0%
H}
{Harris} W.~E.,  1996, VizieR Online Data Catalog, 7195, 0

\bibitem[\protect\citeauthoryear{{Heggie} \& {Mathieu}}{{Heggie} \&
  {Mathieu}}{1986}]{1986LNP...267..233H}
{Heggie} D.~C.,  {Mathieu} R.~D.,  1986, LNP Vol.~267: The Use of
  Supercomputers in Stellar Dynamics, 267, 233

\bibitem[\protect\citeauthoryear{{Holtzman}, {Faber}, {Shaya}, {Lauer},
  {Groth}, {Hunter}, {Baum}, {Ewald} \& {et al.}}{{Holtzman}
  et~al.}{1992}]{1992AJ....103..691Hmnras}
{Holtzman} J.~A.,  {Faber} S.~M.,  {Shaya} E.~J.,  {Lauer} T.~R.,  {Groth} J.,
  {Hunter} D.~A.,  {Baum} W.~A.,  {Ewald} S.~P.,    {et al.} 1992, \aj, 103,
  691

\bibitem[\protect\citeauthoryear{{Kharchenko}, {Piskunov}, {R{\"o}ser},
  {Schilbach} \& {Scholz}}{{Kharchenko} et~al.}{2005}]{2005A&A...438.1163K}
{Kharchenko} N.~V.,  {Piskunov} A.~E.,  {R{\"o}ser} S.,  {Schilbach} E.,
  {Scholz} R.-D.,  2005, \aap, 438, 1163

\bibitem[\protect\citeauthoryear{{King}}{{King}}{1962}]{1962AJ.....67..471K}
{King} I.,  1962, \aj, 67, 471

\bibitem[\protect\citeauthoryear{{King}}{{King}}{1966}]{1966AJ.....71...64K}
{King} I.~R.,  1966, \aj, 71, 64

\bibitem[\protect\citeauthoryear{{Kranz}, {Slyz} \& {Rix}}{{Kranz}
  et~al.}{2001}]{2001ApJ...562..164K}
{Kranz} T.,  {Slyz} A.,    {Rix} H.-W.,  2001, \apj, 562, 164

\bibitem[\protect\citeauthoryear{{Kundic} \& {Ostriker}}{{Kundic} \&
  {Ostriker}}{1995}]{1995ApJ...438..702K}
{Kundic} T.,  {Ostriker} J.~P.,  1995, \apj, 438, 702

\bibitem[\protect\citeauthoryear{{Lamers}, {Gieles}, {Bastian}, {Baumgardt},
  {Kharchenko} \& {Portegies Zwart}}{{Lamers}
  et~al.}{2005}]{2005A&A...441..117L}
{Lamers} H.~J.~G.~L.~M.,  {Gieles} M.,  {Bastian} N.,  {Baumgardt} H.,
  {Kharchenko} N.~V.,    {Portegies Zwart} S.,  2005, \aap, 441, 117

\bibitem[\protect\citeauthoryear{{Lamers}, {Gieles} \& {Portegies
  Zwart}}{{Lamers} et~al.}{2005}]{2005A&A...429..173L}
{Lamers} H.~J.~G.~L.~M.,  {Gieles} M.,    {Portegies Zwart} S.~F.,  2005, \aap,
  429, 173

\bibitem[\protect\citeauthoryear{{Larsen}}{{Larsen}}{2004}]{2004A&A...416..537%
L}
{Larsen} S.~S.,  2004, \aap, 416, 537

\bibitem[\protect\citeauthoryear{{Larsen}, {Brodie}, {Elmegreen}, {Efremov},
  {Hodge} \& {Richtler}}{{Larsen} et~al.}{2001}]{2001ApJ...556..801L}
{Larsen} S.~S.,  {Brodie} J.~P.,  {Elmegreen} B.~G.,  {Efremov} Y.~N.,  {Hodge}
  P.~W.,    {Richtler} T.,  2001, \apj, 556, 801

\bibitem[\protect\citeauthoryear{{Levine}, {Blitz} \& {Heiles}}{{Levine}
  et~al.}{2006}]{2006Sci...312.1773L}
{Levine} E.~S.,  {Blitz} L.,    {Heiles} C.,  2006, Science, 312, 1773

\bibitem[\protect\citeauthoryear{{Makino}, {Fukushige}, {Koga} \&
  {Namura}}{{Makino} et~al.}{2003}]{2003PASJ...55.1163M}
{Makino} J.,  {Fukushige} T.,  {Koga} M.,    {Namura} K.,  2003, \pasj, 55,
  1163

\bibitem[\protect\citeauthoryear{{Martos}, {Hernandez}, {Y{\'a}{\~n}ez},
  {Moreno} \& {Pichardo}}{{Martos} et~al.}{2004}]{2004MNRAS.350L..47M}
{Martos} M.,  {Hernandez} X.,  {Y{\'a}{\~n}ez} M.,  {Moreno} E.,    {Pichardo}
  B.,  2004, \mnras, 350, L47

\bibitem[\protect\citeauthoryear{{McMillan} \& {Hut}}{{McMillan} \&
  {Hut}}{1996}]{1996ApJ...467..348M}
{McMillan} S.~L.~W.,  {Hut} P.,  1996, \apj, 467, 348

\bibitem[\protect\citeauthoryear{{Murali} \& {Weinberg}}{{Murali} \&
  {Weinberg}}{1997}]{1997MNRAS.291..717M}
{Murali} C.,  {Weinberg} M.~D.,  1997, \mnras, 291, 717

\bibitem[\protect\citeauthoryear{{Nakai}, {Kuno}, {Handa} \& {Sofue}}{{Nakai}
  et~al.}{1994}]{1994PASJ...46..527N}
{Nakai} N.,  {Kuno} N.,  {Handa} T.,    {Sofue} Y.,  1994, \pasj, 46, 527

\bibitem[\protect\citeauthoryear{{Nieten}, {Neininger}, {Gu{\'e}lin},
  {Ungerechts}, {Lucas}, {Berkhuijsen}, {Beck} \& {Wielebinski}}{{Nieten}
  et~al.}{2006}]{2006A&A...453..459N}
{Nieten} C.,  {Neininger} N.,  {Gu{\'e}lin} M.,  {Ungerechts} H.,  {Lucas} R.,
  {Berkhuijsen} E.~M.,  {Beck} R.,    {Wielebinski} R.,  2006, \aap, 453, 459

\bibitem[\protect\citeauthoryear{{Oort}}{{Oort}}{1958}]{1958RA......5..507O}
{Oort} J.~H.,  1958, Ricerche Astronomiche, 5, 507

\bibitem[\protect\citeauthoryear{{Ostriker}, {Spitzer} \&
  {Chevalier}}{{Ostriker} et~al.}{1972}]{1972ApJ...176L..51O}
{Ostriker} J.~P.,  {Spitzer} L.~J.,    {Chevalier} R.~A.,  1972, \apjl, 176,
  L51+

\bibitem[\protect\citeauthoryear{{Portegies Zwart}, {McMillan}, {Hut} \&
  {Makino}}{{Portegies Zwart} et~al.}{2001}]{2001MNRAS.321..199P}
{Portegies Zwart} S.,  {McMillan} S.~L.~W.,  {Hut} P.,    {Makino} J.,  2001,
  \mnras, 321, 199

\bibitem[\protect\citeauthoryear{{Rix} \& {Rieke}}{{Rix} \&
  {Rieke}}{1993}]{1993ApJ...418..123R}
{Rix} H.-W.,  {Rieke} M.~J.,  1993, \apj, 418, 123

\bibitem[\protect\citeauthoryear{{Roberts}}{{Roberts}}{1969}]{1969ApJ...158..1%
23R}
{Roberts} W.~W.,  1969, \apj, 158, 123

\bibitem[\protect\citeauthoryear{{Rots}, {Bosma}, {van der Hulst},
  {Athanassoula} \& {Crane}}{{Rots} et~al.}{1990}]{1990AJ....100..387R}
{Rots} A.~H.,  {Bosma} A.,  {van der Hulst} J.~M.,  {Athanassoula} E.,
  {Crane} P.~C.,  1990, \aj, 100, 387

\bibitem[\protect\citeauthoryear{{Schweizer}}{{Schweizer}}{1976}]{1976ApJS...3%
1..313S}
{Schweizer} F.,  1976, \apjs, 31, 313

\bibitem[\protect\citeauthoryear{{Seigar} \& {James}}{{Seigar} \&
  {James}}{1998}]{1998MNRAS.299..685S}
{Seigar} M.~S.,  {James} P.~A.,  1998, \mnras, 299, 685

\bibitem[\protect\citeauthoryear{{Solomon}, {Rivolo}, {Barrett} \&
  {Yahil}}{{Solomon} et~al.}{1987}]{1987ApJ...319..730S}
{Solomon} P.~M.,  {Rivolo} A.~R.,  {Barrett} J.,    {Yahil} A.,  1987, \apj,
  319, 730

\bibitem[\protect\citeauthoryear{{Spitzer}}{{Spitzer}}{1987}]{1987degc.book...%
..S}
{Spitzer} L.,  1987, {Dynamical evolution of globular clusters}.
Princeton, NJ, Princeton University Press, 1987, 191 p.

\bibitem[\protect\citeauthoryear{{Spitzer}}{{Spitzer}}{1958}]{1958ApJ...127...%
17S}
{Spitzer} L.~J.,  1958, \apj, 127, 17

\bibitem[\protect\citeauthoryear{{Spitzer} \& {Chevalier}}{{Spitzer} \&
  {Chevalier}}{1973}]{1973ApJ...183..565S}
{Spitzer} L.~J.,  {Chevalier} R.~A.,  1973, \apj, 183, 565

\bibitem[\protect\citeauthoryear{{Sygnet}, {Tagger}, {Athanassoula} \&
  {Pellat}}{{Sygnet} et~al.}{1988}]{1988MNRAS.232..733S}
{Sygnet} J.~F.,  {Tagger} M.,  {Athanassoula} E.,    {Pellat} R.,  1988,
  \mnras, 232, 733

\bibitem[\protect\citeauthoryear{{Tagger}, {Sygnet}, {Athanassoula} \&
  {Pellat}}{{Tagger} et~al.}{1987}]{1987ApJ...318L..43T}
{Tagger} M.,  {Sygnet} J.~F.,  {Athanassoula} E.,    {Pellat} R.,  1987, \apjl,
  318, L43

\bibitem[\protect\citeauthoryear{{Takahashi} \& {Portegies Zwart}}{{Takahashi}
  \& {Portegies Zwart}}{2000}]{2000ApJ...535..759T}
{Takahashi} K.,  {Portegies Zwart} S.~F.,  2000, \apj, 535, 759

\bibitem[\protect\citeauthoryear{{van der Kruit}}{{van der
  Kruit}}{1988}]{1988A&A...192..117V}
{van der Kruit} P.~C.,  1988, \aap, 192, 117

\bibitem[\protect\citeauthoryear{{van der Kruit} \& {Searle}}{{van der Kruit}
  \& {Searle}}{1981a}]{1981A&A....95..105V}
{van der Kruit} P.~C.,  {Searle} L.,  1981a, \aap, 95, 105

\bibitem[\protect\citeauthoryear{{van der Kruit} \& {Searle}}{{van der Kruit}
  \& {Searle}}{1981b}]{1981A&A....95..116V}
{van der Kruit} P.~C.,  {Searle} L.,  1981b, \aap, 95, 116

\bibitem[\protect\citeauthoryear{{van der Kruit} \& {Searle}}{{van der Kruit}
  \& {Searle}}{1982a}]{1982A&A...110...61V}
{van der Kruit} P.~C.,  {Searle} L.,  1982a, \aap, 110, 61

\bibitem[\protect\citeauthoryear{{van der Kruit} \& {Searle}}{{van der Kruit}
  \& {Searle}}{1982b}]{1982A&A...110...79V}
{van der Kruit} P.~C.,  {Searle} L.,  1982b, \aap, 110, 79

\bibitem[\protect\citeauthoryear{{Weinberg}}{{Weinberg}}{1994a}]{1994AJ....108%
.1398W}
{Weinberg} M.~D.,  1994a, \aj, 108, 1398

\bibitem[\protect\citeauthoryear{{Weinberg}}{{Weinberg}}{1994b}]{1994AJ....108%
.1403W}
{Weinberg} M.~D.,  1994b, \aj, 108, 1403

\bibitem[\protect\citeauthoryear{{Weinberg}}{{Weinberg}}{1994c}]{1994AJ....108%
.1414W}
{Weinberg} M.~D.,  1994c, \aj, 108, 1414

\bibitem[\protect\citeauthoryear{{Whitmore}, {Zhang}, {Leitherer}, {Fall},
  {Schweizer} \& {Miller}}{{Whitmore} et~al.}{1999}]{1999AJ....118.1551W}
{Whitmore} B.~C.,  {Zhang} Q.,  {Leitherer} C.,  {Fall} S.~M.,  {Schweizer} F.,
     {Miller} B.~W.,  1999, \aj, 118, 1551

\bibitem[\protect\citeauthoryear{{Wielen}}{{Wielen}}{1971}]{1971A&A....13..309%
W}
{Wielen} R.,  1971, \aap, 13, 309

\bibitem[\protect\citeauthoryear{{Zimmer}, {Rand} \& {McGraw}}{{Zimmer}
  et~al.}{2004}]{2004ApJ...607..285Z}
{Zimmer} P.,  {Rand} R.~J.,    {McGraw} J.~T.,  2004, \apj, 607, 285

\bibitem[\protect\citeauthoryear{{Zwicky}}{{Zwicky}}{1955}]{1955PASP...67..232%
Z}
{Zwicky} F.,  1955, \pasp, 67, 232

\end{thebibliography}

\end{document}